\documentclass[preprint,12pt]{elsarticle}
\usepackage{amssymb}
\usepackage{amsmath}
\usepackage{amsfonts}
\usepackage{graphicx}
\usepackage{dcolumn}
\usepackage{epstopdf}
\usepackage{bm}
\usepackage{morefloats}

\journal{ADNDT}
\begin{document}
\begin{frontmatter}
\title{Radiative capture of nucleons at astrophysical energies with single-particle states}
\author[tamu,cor]{J.T. Huang}
\ead{jhuang1@leo.tamu-commerce.edu}

\author[tamu]{C. A. Bertulani}
\ead{carlos$\_$bertulani@tamu-commerce.edu}

\author[VG]{V. Guimar\~aes}
\ead{valdirg@dfn.if.usp.br}

\cortext[cor]{Corresponding author}
\address[tamu]{Department of Physics, Texas A\&M University-Commerce, Commerce,
TX 75429, USA}

\address[VG]{Instituto de F\'{\i}sica, Universidade de S\~ao Paulo
P.O.Box 66318, 05389-970 S\~ao Paulo, SP, Brazil}


\begin{abstract}
Radiative capture of nucleons at energies of astrophysical interest
is one of the most important processes for nucleosynthesis. The
nucleon capture can occur either by a compound nucleus reaction or
by a direct  process. The compound reaction cross sections are
usually very small, specially for light nuclei.  The direct capture
proceeds either via the formation of a single-particle resonance, or
a non-resonant capture process. In this work we calculate radiative
capture cross sections and astrophysical S-factors for nuclei in the
mass region $A<20$ using single-particle states. We carefully
discuss the parameter fitting procedure adopted in the simplified
two-body treatment of the capture process. Then we produce a
detailed list of cases for which the model works well. Useful
quantities, such as spectroscopic factors and asymptotic
normalization coefficients, are obtained and compared to published
data.
\end{abstract}

\date{\today}
\end{frontmatter}

\tableofcontents

\listoftables

\listoffigures

\section{Introduction}

Fusion reactions relevant for astrophysics proceed via
compound--nucleus formation, with a very large number of resonances
involved, or by direct capture, with only few or no resonances. To
calculate direct capture cross sections one needs to solve the many
body problem for the bound and continuum states of relevance for the
capture process (for a review see, \cite{RR89}). A much simpler, and
popular, solution is based on a potential model to obtain
single-particle energies and wavefunctions  \cite{Ber03}. The model assumes two
structureless particles interacting via a potential with a relative
coordinate dependence determined by a set of adjusting parameters.
Often, this solution is good enough to yield cross sections
within the accuracy required to reproduce the experiments.

In this article we explore the single-particle model
to perform a systematic
study of radiative capture reactions for several light nuclei.
This study has not yet been reported in the literature, where one finds its
application  to isolated cases. It is
also useful to obtain potential parameters for other reaction channels and
predict quantities of interest, such as spectroscopic factors (SF) and
asymptotic normalization coefficients (ANC).

This paper is organized as follows. In section II we summarize the
theoretical tools used in the single-particle description of
direct capture (DC) reactions. We show how potentials and
wavefunctions are built, followed by a description of how radiative
capture cross sections are obtained. Then we discuss the derivation
and interpretation of the asymptotic normalization coefficients. In
section III we present and discuss the results for radiative proton
capture, whereas in section IV we present and discuss the results
for radiative neutron capture. The sensitivity of the S-factors on the potential
parameters is discussed in section V. A summary of the ANCs obtained in this work
is described in section VI. Our final conclusions are
given in section VII.

\section{Direct capture}

\subsection{Potentials and Wavefunctions}

In this work we adopt nuclear potentials of the form%
\begin{equation}
V(\mathbf{r})=V_{0}(r)+V_{S}(r)\ (\mathbf{l.s})+V_{C}(r) \label{WStot}%
\end{equation}
where $V_{0}(r)$ and $V_{S}(r)$ are the central and spin-orbit interactions,
respectively, and $V_{C}(r)$ is the Coulomb potential of a uniform
distribution of charges:%
\begin{align}
V_{C}(r)  &  =\frac{Z_{a}Z_{b}e^{2}}{r}\ \ \ \text{\ for}\ \ \ \ \ r>R_{C}%
\nonumber\\
&  =\frac{Z_{a}Z_{b}e^{2}}{2R_{C}}\left(  3-\frac{r^{2}}{R_{C}^{2}}\right)
\ \ \ \ \text{for}\ \ \ \ r<R_{C}, \label{coul_pot}%
\end{align}
where $Z_{i}$ is the charge number of nucleus $i=a,b$.

Here we use a Woods-Saxon (WS) parameterization to build up the potentials
$V_{0}(r)$ and $V_{S}(r)$, given by%
\begin{align}
V_{0}(r)  &  =V_{0}\ f_{0}(r),\nonumber\\
V_{S}(r)  &  =-\ V_{S0}\ \left(  \frac{\hbar}{m_{\pi}c}\right)  ^{2}\ \frac
{1}{r}\ \frac{d}{dr}f_{S}(r)\nonumber\\
\text{with}\ \ \ \ f_{i}(r)  &  =\left[  1+\exp\left(  \frac{r-R_{i}}{a_{i}%
}\right)  \right]  ^{-1}\ . \label{centsp}%
\end{align}
The spin-orbit interaction in     Eq. \ref{centsp} is written in terms
of the pion Compton wavelength, $\hbar/m_{\pi}c=1.414$ fm. The
parameters $V_{0}$, $V_{S0}$, $R_{0}$, $a_{0},$ $R_{S0}$, and
$a_{S0}$ are chosen to reproduce the ground state energy $E_{B}$ (or
the energy of an excited state). For this purpose, we define typical
values (Table I) for $V_{S0}$, $R_{0}$, $a_{0},$ $R_{S0}$, and vary
only the depth of the central potential, $V_{0}$. As we discuss
later, a different set of potential depths might be used for
continuum states.

For neutron and proton capture reactions, there is no need for using
another form for the potentials. The WS set of parameters are well
suited to describe any reaction of interest, except perhaps for
those cases in which one of the partners is a neutron-rich halo
nucleus. Then the extended radial dependence leads to
unusual forms for the potentials. Also, for capture reactions in
which the light partner is either a deuteron, tritium, $\alpha
$-particle or a heavier nucleus, folding models are more
appropriate. Folding models are based on an effective
nucleon-nucleon interaction and nuclear densities which are either
obtained experimentally (not really, because only charge densities
can be accurately determined from electron-scattering), or
calculated from some microscopic model (typically Hartree-Fock or
relativistic mean field models). The effective interactions as well
as the nuclear densities are subject of intensive theoretical
studies, which is beyond the scope of this work. We will restrict
our studies to neutron and proton radiative capture reactions based
on a nucleon-nucleus interaction of the form of      Eq. \ref{WStot}.

The wavefunctions for the nucleon (n) + nucleus (x) system are calculated by
solving the radial Schr\"{o}dinger equation%
\begin{equation}
-\frac{\hbar^{2}}{2m_{nx}}\left[  \frac{d^{2}}{dr^{2}}-\frac{l\left(
l+1\right)  }{r^{2}}\right]  u_{\alpha}\left(  r\right)  + V (r) u_{\alpha
}\left(  r\right)  =E_{\alpha}u_{\alpha}\left(  r\right)  \ . \label{bss}%
\end{equation}
The nucleon $n$, the nucleus $x$, and the $n+x=a$--system have
intrinsic spins labeled by $s=1/2$, $I_{x}$ and $J$, respectively.
The orbital angular momentum for the relative motion of $n+x$ is
described by $l$. It is convenient to couple angular momenta as
$\mathbf{l+s}\mathbf{=j}$ and $\mathbf{j+I}_{x}\mathbf{=J}$, where
$\mathbf{J}$ is called the channel spin. In      Eq. \ref{WStot} for $V$
we use $\mathbf{s.l} =\left[ j(j+1)-l(l+1)-3/4\right]  /2$ and
$\alpha$ in      Eq. \ref{bss} denotes the set of quantum numbers,
$\alpha_{b}=\{E_{b},l_{b},j_{b},J_{b}\}$ for the bound state, and
$\alpha_{c}=\{E_{c},l_{c},j_{c},J_{c}\}$ for the continuum states.

The bound-state wavefunctions are normalized to unity, $\int dr \left\vert
u_{\alpha_{b}}\left(  r\right)  \right\vert ^{2}=1$, whereas the continuum
wavefunctions have boundary conditions at infinity given by
\begin{equation}
u_{\alpha_{c}}(r\rightarrow\infty)=i\sqrt{\frac{m_{nx}}{2\pi k\hbar^{2}}%
}\left[  H_{l}^{(-)}\left(  r\right)  -S_{\alpha_{c}}H_{l}^{(+)}\left(
r\right)  \right]  \ e^{i\sigma_{l}\left(  E\right)  } \label{uE}%
\end{equation}
where $S_{\alpha_{c}}=\exp\left[  2i\delta_{\alpha_{c}}\left(
E\right) \right]  $, with $\delta_{\alpha_{c}}\left(  E\right)  $
and $\sigma _{l}\left(  E\right)  $ being the nuclear and the
Coulomb phase-shifts, respectively. In      Eq. \ref{uE},
$H_{l}^{(\pm)}\left(  r\right)  =G_{l}(r)\pm iF_{l}\left(  r\right)
$, where $F_{l}$ and $G_{l}$ are the regular and irregular Coulomb
wavefunctions. For neutrons the Coulomb functions reduce to the
usual spherical Bessel functions, $j_{l}\left(  r\right)  $ and
$n_{l}\left(  r\right)  $. With these definitions, the continuum
wavefunctions are normalized as $\left\langle
u_{E_{c}^{\prime}}|u_{E_{c}} \right\rangle =\delta\left(
E_{c}^{\prime}-E_{c}\right)  \delta_{\alpha\alpha^{\prime}}. $

\subsection{Radiative capture cross sections}

The radiative capture cross sections for $n+x\rightarrow a+\gamma$
and $\pi L$ ($\pi=E,(M)=$electric (magnetic) L-pole) transitions
are calculated with
\begin{align}
\sigma_{EL,J_{b}}^{\text{d.c.}}  & =\frac{(2\pi)^{3}}{k^{2}}\left(
\frac{E_{nx}+E_{b}}{\hbar c}\right)  ^{2L+1}\frac{2(2I_{a}+1)}{(2I_{n}%
+1)(2I_{x}+1)}\nonumber\\
&  \times\frac{L+1}{L[(2L+1)!!]^{2}}\sum_{J_{c}j_{c}l_{c}}(2J_{c}+1)\nonumber\\
&  \times\left\{
\begin{array}
[c]{ccc}%
j_{c} & J_{c} & I_{x}\\
J_{b} & j_{b} & L
\end{array}
\right\}  ^{2}\ \left\vert \left\langle l_cj_c\left\Vert
\mathcal{O}_{\pi L}\right\Vert l_{b}j_{b} \right\rangle
\right\vert^{2},
\label{respf}%
\end{align}
where $E_{b}$ is the binding energy and $\left\langle
l_cj_c\left\Vert \mathcal{O}_{\pi L}\right\Vert l_{b}j_{b}
\right\rangle$ is the multipole matrix element. For the
electric multipole transitions we have
\begin{align}
\left\langle l_cj_c\left\Vert
\mathcal{O}_{EL}\right\Vert l_{b} j_{b}\right\rangle
&=(-1)^{l_b+l_c-j_c+L-1/2}\frac{e_{L}}{\sqrt{4\pi}}
 \nonumber \\
 &\times \sqrt{(2L+1)(2j_b+1)}
\left(
\begin{array}
[c]{ccc}%
j_{b} & L & j_{c}\\
1/2 & 0 & -1/2
\end{array}
\right)
 \nonumber \\
&\times \int_{0}^{\infty}dr \
r^{L}u_{b}(r)u_{c}(r)
,\label{lol0}%
\end{align}
where  $e_{L}$ is the effective charge, which takes into account the
displacement of the center-of-mass,
\begin{equation}
e_{L}=Z_{n}e\left(  -\frac{m_{n}}{m_{a}}\right)  ^{L}+Z_{x}e\left(
\frac{m_{x}}{m_{a}}\right)  ^{L}. \label{el}%
\end{equation}

In comparison with the electric dipole transitions the cross
sections for magnetic dipole transitions are reduced by a factor of
$v^{2}/c^{2}$, where $v$ is the relative velocity of the $n+x$
system. At very low energies, $v\ll c$, $M1$ transitions will be
much smaller than the electric transitions. Only in the case of
sharp resonances, the M1 transitions play a significant role, e.g. for the
$J=1^{+}$ state in $^{8}$B at $E_{R}=630$ keV above the proton
separation threshold \cite{RO73,KPK87}. In general, the potential model
is not good to reproduce M1 transition amplitudes \cite{Bar88}. We will
explore few situations in which the model works well.

The radiative capture cross sections for
$n+x\rightarrow a+\gamma$ and $M1$  transitions are calculated with
\begin{align}
&  \left\langle l_cj_c\left\Vert \mathcal{O}_{M1}\right\Vert l_{b}j_{b}
\right\rangle=\left(  -1\right) ^{j_c+I_{x}+J_{b}+1}\
\sqrt{\frac{3}{4\pi }}\   \mu_{N}\nonumber\\
&  \times\Bigg\{  \frac{1}{\widehat{l}_{b}}e_{M}\Bigg[
\frac{2\widetilde {j}_{b}}{\widehat{l}_{b}}\left(
l_{b}\delta_{j_{b},\ l_{b}+1/2}+
\left( l_{b}+1\right)
\delta_{j_{b},\ l_{b}-1/2}\right)
\nonumber\\
&+\left(  -1\right)
^{l_{b}+1/2-j_c}\frac{\widehat{j}_{b}}{\sqrt{2}}\delta_{j_{b},\
l_{b}\pm
1/2}\delta_{j_c,\ l_{b}\mp1/2}\Bigg]    \nonumber\\
&  +g_{N}\frac{1}{\widehat{l}_{b}^{2}}\Bigg[  \left(  -1\right)
^{l_{b}+1/2-j_{b}}\widetilde{j}_{b}\delta_{j_c,\ j_{b}}
\nonumber \\
&-\left(
-1\right)
^{l_{b}+1/2-j_c}\frac{\widehat{j}_{b}}{\sqrt{2}}\delta_{j_{b},\
l_{b}\pm
1/2}\delta_{j_c,\ l_{b}\mp1/2}\Bigg]  \nonumber\\
&  +g_{x}\left(  -1\right) ^{I_{x}+j_{b}+J_c+1}\widehat{J}_{b}
\widehat{J}_c\widehat{I}_{x}\widetilde{I}_{x}\left\{
\begin{array}
[c]{ccc}%
I_{x} & J_c & j_{b}\\
J_{b} & I_{x} & 1
\end{array}
\right\}  \Bigg\}  \nonumber \\
& \times  \int_{0}^{\infty}dr\  r \ u_{c}\left( r\right) \
u_{b}\left(  r\right)  ,\label{ljolj}
\end{align}
where $\tilde k = \sqrt{k(k+1)}$ and $\hat k = \sqrt{2k+1}$.
The spin g-factor is $g_{N}=5.586$ for the proton and $g_{N}=-3.826$
for the
neutron. The magnetic moment of the core nucleus is given by $\mu_{x}=g_{x}%
\mu_{N}$. If $l_c\neq l_{b}$ the magnetic dipole matrix element is
zero.

The total direct capture cross section is obtained by adding all
multipolarities and final spins of the bound state ($E\equiv E_{nx}$),
\begin{equation}
\sigma^{\text{d.c.}} (E)=\sum_{L,J_{b}} (SF)_{J_{b}}\ \sigma^{\text{d.c.}%
}_{L,J_{b}}(E) \ , \label{SFS}%
\end{equation}
where $(SF)_{J_{b}}$ are spectroscopic factors.

For charged particles the astrophysical S-factor for the direct capture from a
continuum state to the bound state is defined as
\begin{align}
&  S\left(  E\right)  = E\ \sigma^{\text{d.c.}}\left(  E\right)
\;\exp\left[  2\pi\eta\left(  E\right)  \right]  ,\nonumber\\
&  \text{with} \ \ \ \ \ \ \ \eta\left(  E\right)  =Z_{a}Z_{b}e^{2}/\hbar v,
\label{s_lambda}%
\end{align}
where $v$ is the initial relative velocity between $n$ and $x$.

For some resonances, not reproducible with the single-particle model, we will use a simple Breit-Wigner shape
parametrization
\begin{equation}
\sigma_{BW}=\frac{\Gamma}{2\pi}\frac{\sigma_{0}(E)}{(E-E_{R})^{2}+\Gamma
^{2}/4} , \label{BWli8}%
\end{equation}
where $E_{R}$ is the resonance energy. The function $\sigma_{0}(E)$ is given
by
\begin{equation}
\sigma_{0}(E)=\frac{\pi\hbar^{2}}{2m_{xn}E}\frac{2J_{R}+1}{(2Jx+1)(2J_{n}%
+1)}\frac{\Gamma_{n}(E)\Gamma_{\gamma}(E)}{\Gamma(E)}\,
\end{equation}
where the total width $\Gamma=\Gamma_{n}+\Gamma_{\gamma}$ is the sum
of the nucleon-decay and the $\gamma$-decay widths. For simplicity,
and for the cases treated here, we will assume that the resonances
are narrow so that $\sigma_{0}=\sigma(E_R)$.

\subsection{Asymptotic normalization coefficients}

Although the potential model works well for many nuclear reactions of interest
in astrophysics, it is often necessary to pursue a more microscopic approach
\cite{Nav06,QN08} to reproduce experimental data. In a microscopic approach,
instead of the single-particle wavefunctions one often makes use of overlap
integrals, $I_{b}(\mathbf{r})$, and a many-body wavefunction for the relative
motion, $\Psi_{c}(\mathbf{r})$. Both $I_{b}(\mathbf{r})$ and $\Psi
_{c}(\mathbf{r})$ might be very complicated to calculate, depending on how
elaborated the microscopic model is. The variable $\mathbf{r}$ is the relative
coordinate between the nucleon and the nucleus $x$, with all the intrinsic
coordinates of the nucleons in $x$ being integrated out. The direct capture
cross sections are obtained from the calculation of $\sigma_{L,J_{b}%
}^{\text{d.c.}} \propto|\left<  I_{b}(r)||r^{L}Y_{L}|| \Psi_{c}(r)\right>
|^{2}$.

The imprints of many-body effects will eventually disappear at large distances
between the nucleon and the nucleus. One thus expects that the overlap
function asymptotically matches the solution of the Schr\"odinger equation
\ref{bss}, with $V=V_{C}$ for protons and $V=0$ for neutrons. That is, when
$r\rightarrow\infty$,
\begin{align}
I_{b}(r)  &  =C_{1} \frac{W_{-\eta,l_{b}+1/2}(2\kappa r)}{r},
\ \ \ \text{for \ protons}\nonumber\\
&  =C_{2} \sqrt{\frac{2\kappa}{r}}K_{l_{b}+1/2}(\kappa r), \ \ \ \text{for
\ neutrons} \label{whitt}%
\end{align}
where the binding energy of the $n+x$ system is related to $\kappa$ by means
of $E_{b}=\hbar^{2}\kappa^{2}/2m_{nx}$, $W_{p,q}$ is the Whittaker function
and $K_{\mu}$ is the modified Bessel function. In      Eq. \ref{whitt}, $C_{i}$ is
the asymptotic normalization coefficient (ANC).

In the calculation of $\sigma_{L,J_{b}}^{\text{d.c.}}$ above, one often
meets the situation in which only the asymptotic part of $I_{b}(r)$ and
$\Psi_{c}(r)$ contributes significantly to the integral over $r$. In these
situations, $\Psi_{c}(r)$ is also well described by a simple two-body
scattering wave (e.g. Coulomb waves). Therefore the radial integration in
$\sigma_{L,J_{b}}^{\text{d.c.}}$ can be done accurately and the only
remaining information from the many-body physics at short-distances is
contained in the asymptotic normalization coefficient $C_{i}$, i.e.
$\sigma_{L,J_{b}}^{\text{d.c.}}\propto C_{i}^{2}$. We thus run into an
effective theory for radiative capture cross sections, in which the constants
$C_{i}$ carry all the information about the short-distance physics, where the
many-body aspects are relevant. It is worthwhile to mention that these
arguments are reasonable for proton capture at very low energies, because of
the Coulomb barrier.

The spectroscopic factors, $SF$, are usually obtained by adjusting
the calculated cross sections to reproduce the experimental ones.
Here we try to follow the literature as closely as possible. When
experimental data are not available, we use spectroscopic factors
taken from the literature. For the cases in which experimental data
exist, we also try to use spectroscopic factors published in the
literature, and fit the data by varying the depth of the WS
potential for the continuum states.

The asymptotic normalization coefficients, $C_{\alpha}$, can also be
obtained from the analysis of peripheral, transfer and breakup,
reactions. As the overlap integral,      Eq. \ref{whitt}, asymptotically
becomes a Whittaker function, so does the single particle
bound-state wavefunction $u_{\alpha}$, calculated with      Eq.
\ref{bss}. If we call the single particle ANC by $b_{i}$, then the
relation between the ANC obtained from experiment, or a microscopic
model, with the single particle ANC is given by $(SF)_{i}b_{i}^{2}%
=C_{i}^{2}$. This becomes clear from     Eq. \ref{SFS}. The values of
$(SF)_{i}$ and $b_{i}$ obtained with the simple potential model are
useful telltales of the complex short-range many-body physics of
radiative capture reactions. One can also invert this argumentation
and obtain spectroscopic factors if the $C_{i}$ are deduced from a
many-body model, or from experiment, and the $b_{i}$ are calculated
from a single particle potential model \cite{Xu94}.

\section{Proton capture}

Table \ref{t2} summarizes the potential parameters used in cases
where the potential model works reasonably well for
radiative proton capture reactions. A discussion is presented case
by case in the following subsections. Unless otherwise stated, we
use the parameters according to Table \ref{WS} for the single-particle
potential. The parameters for the continuum potential, $V_c$, are
the same as for the bound state potential, except for few
cases discussed explicitly in the text.

\subsection{$\text{d}(\text{p},\gamma)^{3}\text{He}$}

Understanding of the nature of
$^3$He, the only stable 3-body nucleus, constitutes a major advance towards
the solution of the general problem of nuclear forces.
In particular, it involves the influence of the third
nucleon on the interaction between the other two. This
latter interaction has been studied extensively in
deuteron and in nucleon-nucleon scattering.
These are issues beyond the scope of this article. But we will show that a rather
good reproduction of the experimental data for the capture reaction $\text{d}(\text{p},\gamma)^{3}\text{He}$
can be obtained with the simple potential
model described in the previous sections.

The $J_{b}=1/2^{+}$ ground state of $^{3}\text{He}$ is described
as a $j_{b}=s_{1/2}$ proton coupled to the deuterium core, which has
an intrinsic spin $I_{x}=1^{+}$. The gamma-ray transition is
dominated by the $E1$ multipolarity and by incoming $p$ waves. Our
results require a spectroscopic factor $SF=0.7$ to fit the
experimental data shown in Fig. \ref{dp}. If we add d-waves to the
ground-state there is a negligible change in this value. Thus, the
contribution of d-waves in the ground state has been neglected. The
experimental data are from Ref. \cite{GLR62} (filled squares), Ref.
\cite{BKS64} (open squares), Ref. \cite{WBL67} (open circles), Ref.
\cite{SCL95} (filled triangles).

In  Ref. \cite{Mukhamedzhanov95},
the ANC for this reaction was found by an analysis of s-wave pd and
nd scattering. The ANC for the $l=0$ channel was found to be $1.97$
fm$^{-1/2}$ ($C^{2}=3.9\pm0.06$ fm$^{-1}$) \cite{Mukhamedzhanov95}.
Our ANC value is $\sqrt{(SF)b^2}=1.56$ fm$^{-1/2}$, which is in good
agreement with the more complicated analysis presented in  Ref.
\cite{Mukhamedzhanov95}.

\subsection{$^{6}$\textrm{Li}$($\textrm{p}$,\gamma)^{7}\text{Be}$}

Unlike $^7$Li, $^6$Li is predicted
to be formed at a very low level in Big Bang nucleosynthesis,
$^6$Li/H = $10^{-14}$ \cite{Thom93,Van99}.
Whereas most elements are produced by stellar nucleosynthesis, lithium is
mainly destroyed in stellar interiors by thermonuclear reactions
with protons. In fact, $^6$Li is rapidly consumed at stellar
temperatures higher than $2 \times 10^6$ K. The major source of $^6$Li
has been thought for decades to be the interaction of galactic
cosmic rays with the interstellar medium \cite{Nol97}. The low energy capture reaction
$^{6}$\textrm{Li}$($\textrm{p}$,\gamma)^{7}\text{Be}$ plays an important
role in the consumption of $^6$Li and formation of $^7$Be.

The S-factor for this reaction is dominated by captures to the
ground state and the 1st excited state of $^{7}\text{Be}$. Both the
ground state ($J_{b}=3/2^{-}$) and the 1st excited state
($J_{b}=1/2^{-}$) of $^{7}\text{Be}$ are described as a
$j_{b}=p_{1/2}$ proton interacting with the $^{6}$\textrm{Li} core,
which has an intrinsic spin $I_{A}=1^{+}$.  The parameters
calculated according to Table I are used.  The potential depths
which reproduce the ground and excited states are given in Table II.

The continuum state potential depth for transitions to the ground
state is set as $V_{c}=-37.70$ MeV following Ref. \cite{Camargo08}
and the corresponding one for the 1st excited is adjusted to fit the
experimental S-factor for that capture (open circles in Fig.
\ref{6Lip}). In Ref. \cite{Camargo08} the potential parameters and
the spectroscopic factor for the ground state was obtained from a
comparison between a finite-range distorted-wave Born approximation
calculation and the experimental differential cross sections for the
$^9$Be($^8$Li,$^9$Be)$^8$Li elastic-transfer reaction at 27 MeV. The
spectroscopic factors so obtained were compared with shell-model
calculations and other experimental values. The spectroscopic factor
is $0.83$ for the ground state following Ref. \cite{Camargo08} and
$0.84$ for the 1st excited state, following Ref. \cite{Arai02}.

In
Ref. \cite{Arai02}, the reaction is also compared with a calculation
based on a four-cluster microscopic model. The
energy dependence of the astrophysical S-factor for the $^{6}$\textrm{Li}%
$($\textrm{p}$,\gamma)^{7}\text{Be}$ reaction has been studied in
Ref. \cite{Prior04}, as well as in Ref. \cite{Barker80} where an
analysis of the experimental data of Ref. \cite{Swi80} was done. It
was found \cite{Barker80,Prior04} that the gamma-ray transition is
dominated by the $E1$ multipolarity and by incoming $s$ and $d$
waves.

Adopting the spectroscopic values listed above and including $s$ and
$d$ incoming waves, we obtain the result shown in Fig. \ref{6Lip}.
Experimental data are from Ref. \cite{Bruss93} (filled triangles),
Ref. \cite{Switkowski79} (filled squares) and Ref. \cite{Arai02}
(open circles).   The agreement with the experimental data is very
good and consistent with the previous studies
\cite{Swi80,Barker80,Arai02,Camargo08}. Based on these results, we
obtain an ANC ($\sqrt{(SF)b^2}$) of $2.01$ fm$^{-1/2}$ for the
ground state and $1.91$ fm$^{-1/2}$ for the 1st excited state.

\subsection{$^{7}$\textrm{Li}(\textrm{p}, $\gamma$)$^{8}$\textrm{Be}}

The reaction $^{7}$\textrm{Li}(\textrm{p}, $\gamma$)$^{8}$\textrm{Be} is part of the
pp-chain in the Sun, leading to the formation of
$^8$Be \cite{Bah89}. The unstable $^8$Be decays into two $\alpha$-particles in $10^{-16}$ sec.

For this reaction, we consider only the capture to the ground state of $^{8}%
$\textrm{Be} ($J_{b}=0^{+}$), which is described as a
$j_{b}=p_{3/2}$ proton coupled to the $I_{x}=3/2^{-}$
$^{7}$\textrm{Li} core. The gamma-ray transition is dominated by the
$E1$ multipolarity and by incoming $s$ and $d$ waves. In order to
reproduce the resonance  at $0.386$ MeV (in the c.m.), we choose a
spectroscopic factor equal to $0.15$. For the other resonance at
$0.901$ keV (in the c.m.), we chose $SF=0.05$.

The result for both M1 resonances are shown in Figure \ref{7Lip}, by
dashed-dotted curves. The potential depth for the continuum state,
chosen as to reproduce the resonances, are $V_c=-46.35$ MeV and
$V_c=-44.55$ MeV, respectively.  The non-resonant component
(dashed-line) of the S-factor is obtained with $V_{c}=-56.69$ MeV
and $SF=1.0$. The experimental data are from Ref. \cite{Zahnow95}
(open circles). This reaction was also studied  in Ref.
\cite{Sampaio01}. They have obtained an spectroscopic factor of
0.4 for the first M1 resonance at $0.386$ MeV and $SF=1.0$ for the
non-resonant capture. Their analysis is extended to angular
distributions for the capture cross-section and analyzing power at
$E_{p, lab} = 80$ keV which shows a strong E1-M1 interference, which
helps to estimate the spectroscopic amplitudes.

If we only consider the fit to the
non-resonant capture, our ANC ($\sqrt{(SF)b^2}$) is $7.84$
fm$^{-1/2}$. If we choose spectroscopic factors which reproduce the
M1 resonances, the ANC-value evidently changes. This shows that the
ANC extracted from radiative capture reactions with the use of a
potential model are strongly dependent on the presence of
resonances, specially those involving M1 transitions.

\subsection{$^{7}\text{Be}(\text{p},\gamma)^{8}\text{B}$}

The creation destruction of $^7$Be in astrophysical environments is essential for understanding several stellar
and cosmological processes and is not well understood. $^8$B also plays an essential role in understanding our Sun.
High energy $\nu_e$ neutrinos produced by $^8$B decay in
the Sun oscillate into other active species on their way to earth \cite{Ahm01}. Precise predictions of the production
rate of $^8$B solar neutrinos are important for testing solar
models, and for limiting the allowed neutrino mixing parameters.
The  most uncertain reaction leading to $^8$B formation in the Sun is the $^{7}\text{Be}(\text{p},\gamma)^{8}\text{B}$
radiative capture reaction \cite{Junghans03}.

The $J_{b}=2^{+}$ ground state of $^{8}\text{B}$ is described as a
$j_{b}=p_{3/2}$ proton coupled to the $^{7}$\textrm{Be} core, which
has an intrinsic spin $I_{x}=3/2^{-}$. In this case, instead of the
values in Table I, we take $a=0.52$ fm and $V_{so}=-9.8$ MeV. This
is the same set of values adopted in Ref. \cite{Ber03}. The
gamma-ray transition is dominated by the $E1$ multipolarity and by
incoming $s$ and $d$ waves. The spectroscopic factor for
non-resonant transitions is set to $1.0$, which seems to reproduce
best the S-factor for this reaction at low energies. Our results are
shown in Fig. \ref{7Bep}. The experimental data are from Ref.
\cite{VCK70} (open square), Ref. \cite{FED83} (open circles), Ref.
\cite{Baby03,Junghans03,Iwasa99,Kav69} (solid triangle, open
triangle, solid square, solid circle, solid diamond and open
diamond).

In Ref. \cite{Junghans03}, the experimental data is
reproduced with the cluster model calculation of Ref. \cite{DB94}
together with two incoherent Breit-Wigner resonances: a $1^{+}$ M1
resonance at $0.63$ MeV fitted with $\Gamma_{p}=35.7\pm0.6$ keV and
$\Gamma_{\gamma}=25.3\pm1.2$ MeV, and a $3^{+}$ resonance at $2.2$ MeV
fitted with $\Gamma_{p}=350$ keV  and $\Gamma_{\gamma}=150\pm30$
MeV. Our calculated M1 resonance (dashed-dotted line) also
reproduces well the data if we use $V_c=-38.14$ MeV, and $SF=0.7$,
with the other parameters according to Table I.  For the
non-resonant E1 transitions we use $V_c=-41.26$ MeV and $SF=1.0$.
The S-factor at $E=0$, $S_{17}(0)$, is equal to 19.41 eV.b, which is
10\% smaller than that from the most recent experimental and
theoretical analysis \cite{Junghans03,Navratil07}.

A different
experimental approach was used in Ref. \cite{Trache01}, which
extracted the $^8$B ANC from $^8$B breakup reactions at several
energies and different targets. In that reference a slightly lower
value of $S_{17}(0)=16.9\pm 1.7$ eV.b was inferred. That work also
quotes an ANC of $0.67$ fm$^{-1/2}$ ($C^{2}=0.450(30)$ fm$^{-1}$).
Our ANC, extracted from our fit to the radiative capture reaction,
is $\sqrt{(SF)b^2}=0.72$ fm$^{-1/2}$, not much different from  Ref.
\cite{Trache01}.

\subsection{$^{8}$\textrm{B}(\textrm{p}, $\gamma$)$^{9}$\textrm{C}}

Nucleosynthesis of light nuclei is hindered by the
gaps at $A=5$ and $A=8$.
The gap at $A=8$ may be bridged by reactions involving the
unstable nuclei $^8$Li ($T_{1/2}=5840$ ms) and $^8$B ($T_{1/2}=
5770$ ms).
The $^8$B(p,$\gamma$)$^9$C reaction breaks out to a hot part of the pp
chain at temperatures such that this reaction becomes faster than the
competing $\beta^+$ decay. This reaction is especially
relevant in low-metallicity stars with high masses
where it can be faster than the
triple-$\alpha$ process. It is also important under nova conditions. In
both astrophysical scenarios this happens at temperatures
several times larger than $10^8$ K, corresponding to
Gamow window energies around $E=50 - 300$ keV \cite{Wie89,Ful86,Bof93}.

The capture process for this reaction is dominated by E1 transitions
from incoming s waves to bound p states \cite{Mohr03} and the
present work is restricted to an analysis of the capture to the
ground state of $^{9}$\textrm{C }($J_{b}=3/2^{-}$), which is
described as a $j_{b}=p_{3/2}$ proton coupled to the
$^{8}$\textrm{B} core, which has an intrinsic spin $I_{x}=2^{+}$.
The spectroscopic factor has been set to $1.0$ as in Ref.
\cite{Mohr03}, where several spectroscopic factor values are
compared.

A renormalized folding potential for the continuum state
is used in Ref. \cite{Mohr03}, while in our calculation $V_{c}$ is
adjusted to $-22.55$ MeV to yield a similar result. This is done because
there are no experimental data for this reaction.
The results of both calculations are shown in
Fig. \ref{8Bp}. The open circle at $E=0$ is from Refs.
\cite{Trache02, Beaumel01}, which is an extrapolated value from  a
potential model using an ANC deduced from a breakup experiment. Ref.
\cite{Mohr03} also generates resonances by changing parameters of the folding potential.
The ANC found in Ref.
\cite{Guo05} is $1.15$ fm$^{-1/2}$ ($C^{2}=1.33\pm0.33$ fm$^{-1}$), whereas
our ANC ($\sqrt{(SF)b^2}$) = $1.31$ fm$^{-1/2}$.

\subsection{$^{9}\text{Be}(\text{p},\gamma)^{10}\text{B}$}

The reaction $^9$Be(p,$\gamma$ )$^{10}$B plays an important role in primordial
and stellar nucleosynthesis of light elements in the p-shell \cite{RR89,Fow84}.
Hydrogen burning in second generation stars occurs via the  proton-proton (pp) chain and CNO-cycle, with the
 $^9$Be(p,$\gamma$ )$^{10}$B  reaction serving as an intermediate link between these cycles

The $J_{b}=3^{+}$ ground state of $^{10}\text{B}$ is described as
a $j_{b}=p_{3/2}$ proton coupled to the $^{9}$\textrm{Be} core,
which has an intrinsic spin $I_{A}=3/2^{-}$. The gamma-ray
transition for the DC is dominated the $E1$ multipolarity and by
incoming $s$ waves. A spectroscopic factor $SF=1.0$ is used, which is the same value
adopted in Ref. \cite{Miura92}. This value reproduces $^9$Be(d,n)$^{10}$B and $^9$Be($^3$He,d)$^{10}$B  reactions
at incident energies of $10-20$ MeV, and $^9$Be($\alpha$,t)$^{10}$B at 65 MeV. It is also in accordance with
the theoretical predictions of Refs. \cite{CK67,VG69}.

The potential depth for the continuum
state $V_{c}=-31.82$ MeV has been adjusted so that we can
reproduce the direct capture measurements reported in Refs. \cite{ZAJ95}. It  also reproduces the
results of
Ref. \cite{Sattarov99}  where a reanalysis of the existing
experimental data on $^{9}\text{Be}(\text{p},\gamma)^{10}\text{B}$ was done within the framework of the R-matrix method.
The direct capture part of the S-factor was calculated using the experimentally measured ANC for $^{10}\text{B} \rightarrow ^9\text{Be}$
+ p. The results are shown in  Fig.
\ref{11Bep}. The experimental data are from Ref. \cite{ZAJ95}
(filled squares). These data have also been fitted in Ref.
\cite{Barker02} using R-matrix formulae that include channel
contributions where appropriate. The curve passing through the experimental
data points is the sum of our DC calculation and the resonance fits, given by the dashed lines.

In Ref. \cite{Mukhamedzhanov97}, the differential cross sections for the reactions $^9$Be($^{10}$B,$^{10}$B)$^9$Be and $^9$Be($^{10}$B,$^9$Be)$^{10}$B have been measured
at an incident energy of 100 MeV. By normalizing the theoretical  cross sections to the experimental data, the ANC for $^{10}\text{B} \rightarrow ^9\text{Be}$ + p was extracted
and found to be $2.22$ fm$^{-1/2}$ ($C^{2}=4.91$ fm$^{-1}%
$), whereas the ANC ($\sqrt{(SF)b^2}$) obtained from our fit to the
previous analysis of Refs. \cite{Miura92,Barker02}  is $3.43$
fm$^{-1/2}$.

\subsection{$^{11}$\textrm{C}(\textrm{p}, $\gamma$)$^{12}$\textrm{N}}

For first generation stars (those composed entirely of nuclei produced in the Big Bang) can
only undergo nucleosynthesis via the pp chains or the triple-alpha process
until heavier nuclei are produced to initiate the CNO cycle. For supermassive first generation stars,
such processes generate too little energy and the triple-alpha reaction turns on too late to cause
an explosion. Rather, such stars could simply collapse to black holes. However, hot pp chains
provide a path for supermassive first generation stars to produce CNO nuclei at a lower temperature than
required by the triple-alpha reaction \cite{Wie89}. These CNO nuclei then serve as seeds for
further energy generation, stablizing the star against collapse long enough to permit an
explosion to occur.
Both the $^8$B(p,$\gamma$)$^9$C and $^{11}$C(p,$\gamma$)$^{12}$N reactions are important in the hot pp chains.

For the $^{11}$\textrm{C} (\textrm{p}, $\gamma$)$^{12}$\textrm{N}
reaction, non-resonant capture into the ground state of
$^{12}$\textrm{N} and resonant capture into the first and second
excited states is thought to dominate the reaction rate at stellar energies
\cite{Tang03}. There are no experimental data for this reaction, except for
indirect determination of the astrophysical S-factors, e.g. by using the ANC for $^{12}$N $\rightarrow ^{11}$C + p from
the $^{14}$N($^{11}$C, $^{12}$N)$^{13}$C peripheral transfer reaction.
Another indirect measurement for the astrophysical rate of the $^{11}$C + p reaction was obtained from  from the
Coulomb break-up of a $^{12}$N radioactive beam in Ref. \cite{Lefebvre95}.

The ground state of $^{12}$\textrm{N} ($J_{b}=1^{+}$)
is described as a $j_{b}=p_{1/2}$ proton coupled to the
$^{11}$\textrm{C} core, which has an intrinsic spin $I_{x}=3/2^{-}$.
The direct capture gamma-ray transition is dominated by the $E1$
multipolarity and by incoming $s$ waves. The spectroscopic factor
has been set to $0.4$, the same value used in Ref. \cite{Lefebvre95} ($0.4\pm 0.25$).

The result for our DC calculation is shown in Fig. \ref{11Cp}.
Similar results have also been reported in Refs.
\cite{Tang03,Lefebvre95, Tribble03}, and in Ref. \cite{Liu03} which
also used the ANC method to extract the low-nergy S-factor via
measurement of $^{11}$C(d, n)$^{12}$N reaction. The ANC found in
Ref. \cite{Tang03} is $1.32$ fm$^{-1/2}$ ($C^{2}=1.73\pm0.25$
fm$^{-1}$) and in Ref. \cite{Liu03}\ is $1.69$ fm$^{-1}$
($C^{2}=2.86\pm0.91$ fm$^{-1}$). Our  ANC value is $\sqrt{(SF)b^2} =
0.94$ fm$^{-1/2}$.

\subsection{$^{12}$\textrm{C}(\textrm{p}, $\gamma$)$^{13}$\textrm{N}}
The abundance ratio $^{12}$C/$^{13}$C is an important measure of stellar evolution
and nucleosynthesis.   Changes in ratios of  $^{12}$C/$^{13}$C
in stars happen as they evolve from the main sequence to the
first ascent giant branch. Later, the convection zone
grows and penetrates to greater depths where it begins
to dredge up material that has been hot enough for the
CNO cycle to convert to N. This is when the primordial
$^{12}$C is converted into $^{13}$C and $^{14}$N by the reactions
$^{12}$\textrm{C}(\textrm{p}, $\gamma$)$^{13}$\textrm{N}($\beta^+$)$^{13}$C and $^{13}$C(p,$\gamma$ )$^{14}$N,
hence reducing the $^{12}$C/$^{13}$C ratio. During the late asymptotic giant branch
AGB phase, the stars suffer thermal instabilities in the helium
shell where partial helium burning occurs causing the
$^{12}$C/$^{13}$C ratio to increase \cite{RR89}.

The {$^{12}$\textrm{C}(\textrm{p}, $\gamma$)$^{13}$\textrm{N}  direct capture to the ground state
proceeds primarily through E1 ($s_{1/2}\rightarrow p_{1/2}$) and E1 ($d_{3/2}%
\rightarrow p_{1/2}$) single-particle transitions \cite{Nesaraja01}.
The ground state of $^{13}$\textrm{N} ($J_{b}=1/2^{-}$) is described
as a $j_{b}=p_{1/2}$ proton coupled to the $^{12}$\textrm{C} core,
which has an intrinsic spin $I_{x}=0^{+}$.

Experimental cross sections for the $^{12}$C(p,$\gamma$) capture to
the ground state of $^{13}$\textrm{N} were published in  Ref.
\cite{Rolfs74}. Choosing the spectroscopic factor as $SF=1$, leads
to the dashed line shown in Fig. \ref{12Cp}, if we use the same
potential depth as for the bound state. The E1 resonance at 0.422
MeV \cite{Nesaraja01} is  generated when we choose $V_{c}= -53.50$
MeV and a spectroscopic factor equal to 0.35. The result for the
resonance is shown as a dotted-line in Fig. \ref{12Cp}. The total
S-factor is shown by a solid line.

The resonance at 0.422 MeV (c.m.) has also been studied
experimentally and theoretically in Refs. \cite{Langanke85,
Burtebaev08, Mahaux65}. The ANC obtained in Refs.
\cite{Yarmukhamedov97, Artemov02, Burtebaev08} is $1.43\pm0.09$
fm$^{-1/2}$, whereas our ANC ($\sqrt{(SF)b^2}$), corresponding to
the non-resonant capture, is $2.05$ fm$^{-1/2}$.

\subsection{$^{13}$\textrm{C}(\textrm{p}, $\gamma$)$^{14}$\textrm{N}}

This  reaction is another important reaction in the CNO cycle. It
precedes the slowest reaction in the CNO cycle, the $^{14}$N(p, $\gamma$ )$^{15}$O radiative capture
reaction, which defines the rate of energy production in the cycle. The
$^{13}$\textrm{C}(\textrm{p}, $\gamma$)$^{14}$\textrm{N} radiative capture rate is also important for
nucleosynthesis via the slow proton capture process
because it depletes the seed nuclei required for the neutron generator reaction
$^{13}$C($\alpha$, n)$^{16}$O in AGB stars with solar metalicity \cite{Ulr82,Lug03}.

Extensive experimental data on this reaction was published in  Ref. \cite{King94}. One concludes that
this capture is dominated by transitions to the ground state. The
direct capture to the ground state proceeds primarily through E1
($s_{1/2}\rightarrow p_{1/2}$) and E1 ($d_{3/2}\rightarrow p_{1/2}$)
single-particle transitions \cite{Nesaraja01}. The ground state of
$^{14}$\textrm{N} ($J_{b}=1^{+}$) is described as a $j_{b}=p_{1/2}$
proton coupled to the $^{13}$\textrm{C} core, which has an intrinsic
spin $I_{x}=1/2^{-}$.

We could not reproduce  the E1 resonance at $E_{cm}=0.518$ MeV using the potential
parameters of Ref. \cite{Nesaraja01}. In fact, we notice that figure 5 of Ref. \cite{Nesaraja01} is inconsistent with its caption
(center of mass and laboratory systems are swapped).
In figure  \ref{13Cp}, the dotted line is our calculation for the resonance, which is obtained with the parameters from Table I and with  generated when
$V_{c}=-50.60$ MeV and spectroscopic factor 0.15. For non-resonant capture, the potential depth for the continuum state
has been chosen as $V_{c}=-44.10$  MeV to reproduce the same result as in Ref. \cite{Nesaraja01}. The spectroscopic factor has
been set to $0.33$ following Ref. \cite{King94}. The non-resonant calculation is
shown by a dashed line in Fig. \ref{13Cp}. The total
S-factor is shown as a solid line.

In Ref. \cite{Mukhamedzhanov03NPA}, the $^{13}$C(p, $\gamma$
)$^{14}$N radiative capture reaction is analyzed within the R-matrix
approach. The experimental ANCs induced from the $^{13}$C($^{14}$N,
$^{13}$C)$^{14}$N and $^{13}$C($^3$He, d)$^{14}$N reactions are used
in the analysis. The obtained ANC is $4.3$ fm$^{-1/2}$ ($C^{2}=18.2$
fm$^{-1}$), whereas our  ANC value is $\sqrt{(SF)b^2} = 3.05$
fm$^{-1/2}$.

\subsection{$^{13}$\textrm{N}(\textrm{p},$\gamma$)$^{14}$\textrm{O}}

For temperatures up to $10^9$ K ($T_9=1$), this reaction is vital for understanding hydrogen burning in the
hot CNO cycle and the conditions under which break-out into the rp-process might occur.

The ground state of $^{14}$\textrm{O} ($J_{b}=0^{+}$) is described
as a $j_{b}=p_{1/2}$ proton coupled to the $^{13}$\textrm{N} core,
which has an intrinsic spin $I_{x}=1/2^{-}$. The gamma-ray
transition for the DC  to the ground state is dominated the $E1$
multipolarity and by incoming $s$ waves.
For the non-resonant capture (lower curves in figure \ref{13Np}),
if we choose the potential depth for
the continuum state to be the same as that for the bound state
($V_{c}=V_{b}$), we obtain S-factors which are up to 3 times larger than the one in
Ref. \cite{Bing07} where a theoretical description of this reaction has been reported. We find that $V_{c}=-25.20$ MeV reproduces rather well the non-resonant
capture calculation of Ref. \cite{Bing07}.
The spectroscopic factor has been set to $1.88$ following Ref.
\cite{Li06}, where a DWBA analysis of the $^{13}$N(d, n)$^{14}$O
reaction at $E_{cm} = 8.9$ MeV was used to obtain the ANC
for the ground state of $^{14}$O$\longrightarrow^{13}$N + p. Our non-resonant DC calculation is shown as a dotted-dashed line in Fig.
\ref{13Np}.

We reproduce the E1 resonance at 0.528 MeV (s wave capture,
according to Ref. \cite{Tang04}) as shown by the solid line in
figure \ref{13Np} if  we choose $V_c=-52.14$ MeV and $SF=1.0$. Very
similar results were obtained in Refs. \cite{Descouvemont89,
Decrock93, Delbar93, Tang04, Li06, Langanke85}. The ANCs reported in
there publications are  $5.51$ fm$^{-1/2}$ ($C^{2}=30.4\pm7.1$
fm$^{-1}$)  \cite{Bing07}, $5.42\pm0.48$ fm$^{-1/2}$  \cite{Li06}
and $5.39$ fm$^{-1/2}$ ($C^{2}=29.0\pm4.3$ fm$^{-1}$) \cite{Tang04}.
In Ref. \cite{Tang04} the radiative capture cross section was
estimated using an R-matrix approach with the measured ANC  from the
$^{14}$N($^{13}$N,$^{14}$O)$^{13}$C peripheral transfer reaction at
11.8 MeV/nucleon incident energy. We obtain the ANC value
$\sqrt{(SF)b^2} = 5.44$ fm$^{-1/2}$, which is in accordance with
these results.

\subsection{$^{14}\text{N}$($\text{p}$,$\gamma$)$^{15}$\textrm{O}}

At astrophysical energies, this reaction is the slowest process in the hydrogen burning CNO cycle \cite{RR89}.
It plays a key role in the energy production of more massive main
sequence stars and the detailed understanding of the neutrino
spectrum of our sun \cite{Ahl90,Arp08} as well as the age determination of
globular cluster stars \cite{Imb05}.

The main contribution to the S-factor for this reaction is due to the
transition to the $6.793$ MeV excited state of $^{15}\text{O}$
($J_{b}=3/2^{+}$) \cite{Formicola04, Schurmann08}, which is
described as a $j_{b}=s_{1/2}$ proton coupled to the
$^{14}$\textrm{N} core ($I_{x}=1^{+}$). The gamma-ray transition is
dominated the $E1$ multipolarity and by incoming $p$ waves. In this
case, instead of the parameters of Table \ref{WS}, we use $r_{0}=1.3$ fm, $r_{c}=1.2$, $a=0.6$ fm and $V_{s0}=-2.0$ MeV,
which were also adopted in Ref. \cite{Nelson03}. The spectroscopic factor is
$SF=1.5$. The result of our calculation is shown by the dashed line in Fig.
\ref{14Np}.

Ref. \cite{Raman94} discusses experiments on stripping reactions and theoretical
shell-model calculations used to describe the $^{14}$\textrm{N}%
$({\rm p},\gamma)^{15}\text{O}$ radiative capture.
These studies indicate that the resonance at
$0.259$ MeV decays to the $6.793$ MeV excited state of
$^{15}$\textrm{O} via a M1 radiation. To describe this resonance, we use the
same spectroscopic factor, $SF=0.51$, as obtained experimentally in Ref.
\cite{Bertone02} where spectroscopic factors and ANCs have been determined for bound
states in $^{15}$O using the $^{14}$N($^3$He,d)$^{15}$O reaction. Several other spectroscopic values from the literature were also
discussed in Ref. \cite{Bertone02}. We found an  optimal value for the potential depth to be $V_c=-49.18$ MeV. Adopting this value, our
calculations yield the dotted line shown in the figure.

The total S-factor is shown as a solid line. Experimental data are
from Ref. \cite{Formicola04} (filled squares), Ref. \cite{Pixley57}
(filled triangles), Ref. \cite{Schroeder87} (open circles). The
R-matrix method was used to study this reaction in Ref.
\cite{Formicola04} and Ref. \cite{Anguloa05}. The ANC found in Ref.
\cite{Mukhamedzhanov03}, deduced from the $^{13}$C($^{14}$N,
$^{13}$C)$^{14}$N and $^{13}$C($^3$He, d)$^{14}$N reactions, is
$4.90$ fm$^{-1/2}$ ($C^{2}=24\pm5$ fm$^{-1}$). Ref. \cite{Bertone02}
adopts the value $4.6$ fm$^{-1/2}$ ($C^{2}=21\pm5$ fm$^{-1}$). Our
ANC obtained from the DC fitting is $\sqrt{(SF)b^2}=5.19$
fm$^{-1/2}$.

\subsection{$^{15}$\textrm{N}$(\text{p},\gamma)^{16}\text{O}$}

In second-generation stars with masses larger than the mass of the
Sun, hydrogen burning proceeds predominantly through the CNO cycle
\cite{RR89}. The  main sequence of reaction leads to an energy
release of 25 MeV per cycle. There is a loss of CN catalyst from
this cycle through the reaction $^{15}$N(p, $\gamma$)$^{16}$O. This
is replenished by a sequence of reactions involving oxygen and
fluorine, leading to the formation of $^{14}$N and $^{15}$N. The
reaction rate of $^{15}$N(p, $\gamma$)$^{16}$O determines the
overall abundance of the oxygen isotopes synthesized in the CNO
tri-cycle \cite{RR89} and therefore plays an important role in
stellar nucleosynthesis.

This reaction is dominated by the capture into the ground state of $^{16}%
$\textrm{O }($J_{b}=0^{+}$) \cite{RR74}, which is described as a
$j_{b}=p_{1/2}$ proton coupled to the $^{15}$\textrm{N} core ($I_{x}=1/2^{-}%
$).  The reaction is dominated by resonant capture to the ground state through
the first two interfering $J^\pi = 1^-$ s-wave resonances at $E_{cm} = 312$ and
964 keV

We will restrict ourselves to the non-resonant capture to the ground state, as a good
reproduction of the resonances is not possible with the simple potential model.
The non-resonant
capture process proceeds predominantly through an E1 ($s\rightarrow
p$) transition \cite{RR74}.
A spectroscopic factor $SF=1.8$ is used for the ground state of
$^{16}$\textrm{O}, following Ref. \cite{RR74} which studied
the excitation functions of this reaction at $E_p = 150-2500$ keV. This value is also
in accordance with Ref. \cite{Ful69}.

Our calculation is shown in Fig.
\ref{15Np}. Experimental data are from Ref. \cite{RR74} (filled
triangles), Ref. \cite{Hebbard60} (filled squares). Ref. \cite{MBB08} extracted ANCs from
the differential cross sections for the $^{15}$N($^3$He,d)$^{16}$O reaction.
Using these ANCs and proton and $\alpha$-resonance widths determined from an R-matrix fit to
the data from
the $^{15}$N(p, $\alpha$)$^{12}$C reaction, the astrophysical factor for  $^{15}$\textrm{N}%
$({\rm p},\gamma)^{16}\text{O}$ was obtained.   The results from Ref.
\cite{RR74} and Ref. \cite{MBB08} are also shown in Fig. \ref{15Np}.
In Ref. \cite{RR74}, the resonances are described by  using a fit
with single level Breit-Wigner shapes. The ANC found in Ref.
\cite{MBB08} is $13.86$ fm$^{-1/2}$ ($C^{2}=192.0\pm26.0$
fm$^{-1}$). Our ANC is very close to this value, i.e.,
$\sqrt{(SF)b^2} = 13.63$ fm$^{-1/2}$.

\subsection{$^{16}\text{O}(\text{p},\gamma)^{17}\text{F}$}

Many stars, including the Sun, will eventually pass through
an evolutionary phase that is referred to as the asymptotic giant
branch (AGB) \cite{Her05}. This phase involves a
hydrogen and a helium shell that burn alternately surrounding
an inactive stellar core.  The $^{16}\text{O}(\text{p},\gamma)^{17}\text{F}$ reaction rate
influences sensitively the $^{17}$O/$^{16}$O isotopic ratio predicted by
models of massive ($\ge 4M_\odot$) AGB stars, where proton captures
occur at the base of the convective envelope (hot bottom burning).
A fine-tuning of the $^{16}\text{O}(\text{p},\gamma)^{17}\text{F}$  reaction rate may account
for the measured anomalous $^{17}$O/$^{16}$O abundance ratio in small grains
with are formed by the condensation of the material ejected from the surface of
AGB stars via strong stellar winds \cite{Hab04}.

We calculate the capture to the ground state and to the 1st excited state of $^{17}%
\text{F}$. The $J_{b}=5/2^{+}$ ground state
($J_{b}=1/2^{+}$ excited state) $^{17}\text{F}$ is described as a
$j_{b}=d_{5/2}$ neutron ($j_{b}=s_{1/2}$ neutron) coupled to the $^{16}%
$\textrm{O} core, which has an intrinsic spin $I_{x}=0^{+}$. In this
case, the values $a=0.6$ fm and $R_{C}=R=3.27$ fm are adopted, which
are the same values used in Ref. \cite{BD03}. The gamma-ray
transitions are dominated by the $E1$ multipolarity and by incoming
$p$ waves for both states. The M1 and E2 contributions amount to
less than $0.1\%$ of the dominant E1 contribution, as shown in Ref. \cite{Rolfs73}
where a potential model was also used.

We
use spectroscopic factors equal to $0.9$ and $1.0$ for the ground state
and the excited state, respectively, following Ref. \cite{Rolfs73}.
Our results are shown in Fig. \ref{16Op}. The experimental data are
from Ref. \cite{HPL58} (filled squares), Ref. \cite{Tanner59}
(filled triangles), Ref. \cite{Rolfs73} (open circles), and Ref.
\cite{MKM97} (open triangles).

Ref. \cite{Gagliardi99} reports a study of the
$^{16}$O($^3$He,d)$^{17}$F reaction  to determine ANCs for
transitions to the ground and first excited states of $^{17}$F.  The
ANCs found in Ref. \cite{Gagliardi99} are $1.04$ fm$^{-1/2}$
($C^{2}=1.08 \pm 0.1$ fm$^{-1}$) for the ground state, and $80.6$
fm$^{-1/2}$ ($C^{2}=6490\pm 680$ fm$^{-1}$) for the first excited
state of $^{17}$F, respectively. Our ANC values are
$\sqrt{(SF)b^2}=0.91$ fm$^{-1/2}$ for the ground state and $77.21$
fm$^{-1/2}$ for the 1st excited state.

\subsection{$^{20}\text{Ne}$($\text{p}$,$\gamma$)$^{21}\text{Na}$}

Along with the p-p chain and the CNO tri-cycle, the Ne-Na cycle \cite{Mar57}  is also of
importance in hydrogen burning in second-generation stars with masses
larger than the mass of the Sun.
The $^{20}$Ne(p, $\gamma$)$^{21}$Na reaction
is the first reaction of the cycle. The nuclei $^{21}$Na, $^{21}$Ne, $^{22}$Na, $^{22}$Ne, and
$^{23}$Na are gradually created during Ne-Na burning. $^{21}$Ne is of additional
interest for subsequent He burning in stars. Due to the positive Q-value of 2.56 MeV
for the $^{21}$Ne($\alpha$, n)$^{24}$Mg reaction, $^{21}$Ne can act as a source of neutrons. Subsequent
capture of these neutrons contributes to the synthesis of the heavier elements  \cite{Mar57}.

As observed in Ref. \cite{RRSW75}, the direct capture to the $2.425$
MeV ($J^{\pi}=1/2^{+}$) and $0.332$ MeV ($J^{\pi}=5/2^{+}$) excited
state dominate the total S-factor for this reaction. The
$J_{b}=1/2^{+}$ excited state ($J_{b}=5/2^{+}$ excited state) of
$^{21}\text{Na}$ is described as a $j_{b}=s_{1/2}$ proton
($j_{b}=d_{5/2}$ proton) interacting with the $^{20}$\textrm{Ne}
core, which has an intrinsic spin $I_{x}=0^{+}$. The gamma-ray
transition is dominated by the $E1$ multipolarity and by incoming
$p$ waves.

The spectroscopic factor obtained in Ref. \cite{RRSW75} is $0.9$.
More recently, Ref. \cite{Mukhamedzhanov06} determined the ANC for
$^{21}$Na$ \rightarrow ^{20}$Ne + p from the analysis of
$^{20}$Ne($^3$He, d)$^{21}$Na proton transfer reaction at an
incident energy of 25.83 MeV, and obtained the spectroscopic factor
of 0.6. We used the spectroscopic factor $SF=0.7$ for the $2.425$
MeV excited state and $SF=0.8$ for the $0.332$ MeV excited state,
which are values between those of Refs. \cite{RRSW75} and
\cite{Mukhamedzhanov06}. Our results are shown in Fig. \ref{20Nep}.
Experimental data are from Ref. \cite{RRSW75}.

For the $2.425$ MeV excited stated, the ANC found in Ref.
\cite{Mukhamedzhanov06} is $8.29\times10^{16}$ fm$^{-1/2}$
($C^{2}=6.8694\times10^{33}$ fm$^{-1}$), whereas our computed ANC
value is $\sqrt{(SF)b^2} = 3.36$ fm$^{-1/2}$. The reason for this
large discrepancy is not clear. It might be, as seen from Fig.
\ref{20Nep}, due to the steep slope of the S-factor at low energies.
This points to a subthreshold resonance and a possible large
sensitivity of the ANC for this state. On the other hand, for the $0.332$ MeV excited
state, the ANC found in Ref. \cite{Mukhamedzhanov06} is $1.55$
fm$^{-1/2}$ ($C^{2}=2.41$ fm$^{-1}$), whereas our computed ANC value
is $\sqrt{(SF)b^2} = 2.17$ fm$^{-1/2}$.

\section{Neutron capture}

Table \ref{t3} summarizes the potential parameters used in the cases
where the single-particle model works reasonably well in calculating
radiative neutron capture reactions. A discussion is presented case
by case in the following subsections. Unless otherwise stated, we
use the parameters described in Table \ref{WS} for the single-particle
potential. The parameters for the continuum potential, $V_c$, are
the same as those for the bound state potential, except for the few
cases explicitly discussed in the text.

\subsection{$^{2}$\textrm{H}$(\text{n},\gamma)^{3}\text{H}$}

The $^2$H(p,$\gamma$)$^3$He reaction at low energies, followed by d($^3$He,p)$^4$He,
leads to the formation of  $^4$He  during the
primordial nucleosynthesis era \cite{Pee66,Sat67,Wag67}.
It  also plays a key role during the proto-stars era, in which the energy generated
by deuterium burning slowed down the contraction due to the gravitational force
\cite{Sta88,Cas02}.
On the other hand, the $^2$H(n,$\gamma$)$^3$H reaction is thought
to contribute to  inhomogeneous
big-bang models \cite{App85,Mal88,App88}. These models assume the existence of  neutron-rich
and neutron-poor regions resulting from a first-order phase
transition from quarks to hadrons
as the universe cooled down \cite{App85}. In the neutron-rich region, reactions such as
$^2$H(n,$\gamma$)$^3$H(d,n)$^4$He($^3$H,$\gamma$)$^7$Li(n,$\gamma$)$^8$Li($\alpha$,n)$^{11}$B(n,$\gamma$)$^{12}$B,
produce an appreciable amount of intermediate-heavy nuclei.

We consider only the E1 capture to the ground state of
$^{3}$\textrm{H} ($p\rightarrow s$). The $J_{b}=1/2^{+}$ ground
state $^{3}\text{He}$ is described as a $j_{b}=s_{1/2}$ neutron
coupled to the $^{2}$\textrm{H} core, which has an intrinsic spin
$I_{x}=1^{+}$.

The calculation for this reaction requires a three-body treatment which is beyond the
scope of this work. Obviously, the potential model adopted here is oversimplified for
this case. We choose an spectroscopic factor equal to $SF=1.0$.
Our results are shown in Fig. \ref{dn}, where the dashed and dash-dotted curves are
the evaluated reaction rates presented in Ref. \cite{Nagai06} based on a phenomenological
parametrization of the cross section based on evaluated nuclear data tables.  The experimental
data are from Ref. \cite{Nagai06}. In Ref. \cite{Nagai06} the
neutron-deuteron capture was obtained using time reversal from the two-body photodisintegration amplitude
and wavefunctions obtained with  the AV18 potential \cite{Wir95} alone
or combined with the Urbana IX three-nucleon force \cite{Pud97}.
Their results are shown by the open (solid) triangles with (without) the three-body nn interaction.
The ANC calculated with our potential model is $\sqrt{(SF)b^2}=1.90$
fm$^{-1/2}$.

\subsection{$^{7}$$\text{Li}$($\text{n}$,$\gamma$)$^{8}$$\text{Li}$}

The $^7$Li(n,$\gamma$)$^8$Li cross section is often used to extrapolate the
capture cross section for the reaction $^7$Be(p,$\gamma$)$^8$B down to the solar energies
at $E_{cm} \sim 20$ keV, which is relevant for the production of high energy neutrinos in the
Sun \cite{Fil83}. The $^7$Li(n,$\gamma$)$^8$Li reaction is also relevant for the rapid process during
primordial nucleosynthesis of nuclei with $A>12$ in the inhomogeneous big-bang models  \cite{App88,Ful88}. In
these models, the main reaction chain leading to the
synthesis of heavy elements is \cite{App88} $^1$H(n,$\gamma$)$^2$H(n,$\gamma$)$^3$H(d,n)$^4$He(t, $\gamma$)$^7$Li(n, $\gamma$)$^8$Li,
and then $^8$Li($\alpha$,n)$^{11}$B (n,$\gamma$)$^{12}$B($\beta^-$)$^{12}$C(n,$\gamma$)$^{13}$C, etc., for heavier nuclei.
The reaction $^7$Li(n,$\gamma$)$^8$Li  is thus a crucial input to bridge the gap of mass $A = 8$, leading to the production of
heavy elements.

We consider the capture to the ground state and to the first excited state of $^{8}%
$\textrm{Li}. A similar calculation has been done in Ref.
\cite{Nagai}, where the partial cross sections from neutron capture
to the ground and first excited states in $^{8}$\textrm{Li} at
stellar energies were reported. The gamma-ray transitions are
dominated by the $E1$ multipolarity and by incoming $s$ waves and
$d$ waves. The $J_{b}=2^{+}$ ground state ($J_{b}=1^{+}$ first
excited state) of $^{8}\text{Li}$ is described as a $j_{b}=p_{3/2}$
neutron interacting with the $^{7}\text{Li}$ core, which has an
intrinsic spin $I_{x}=3/2^{-}$.

In this particular case, the values $R_0=R_C=R_{S0}=2.391$ fm are
used. For the continuum state, the potential depth has been adjusted
to reproduce the experimental scattering lengths
$a_{+}=-3.63\pm0.05$ fm and $a_{-}=+0.87\pm0.05$ fm for the two
components of the channel spin $s$ at thermal energies. The
resulting potential depth parameters are $V_{c}=-56.15$ MeV and
$V_{c}=-46.50$ MeV, for the $s=2$ and $s=1$ spin components,
respectively. Following Ref.  \cite{Nagai}, we use the spectroscopic
factors $SF(g.s.)=0.87$ and $SF(1st)=0.48$, for the ground and first
excited states, respectively. The capture to the first excited state
contributes to less than 5\% of the total cross section. The M1
resonance at $E_{R}=0.26$ MeV for capture to the ground state is
reproduced with $V_{c}=-34.93$ MeV and a spectroscopic factor
$SF=1.0$.

The results of this calculation are shown in  Fig. \ref{7Lin}. The
dashed and dotted lines
are for the capture to the ground state and first excited state,
respectively. Adding them together with the dashed-dotted line for the M1 resonance, one gets the total S-factor shown by the
solid line. The experimental data are from refs. \cite{Imhof}
(filled circles), \cite{Nagai} (filled triangles), \cite{WSK89}
(filled squares), \cite{Nag91} (open circles) and \cite{Hei98} (open
triangles). Our calculated ANC is $\sqrt{(SF)b^2}= 0.71$
fm$^{-1/2}$ for the ground state  and $0.33$
fm$^{-1/2}$ for the 1st excited state of $^{8}$\textrm{Li}.

\subsection{$^{8}\text{Li}(\text{n},\gamma)^{9}\text{Li}$}

Rapid capture processes (r-processes) might occur in the post-collapse of a type II supernova,
leading to the formation of heavy elements.
Starting with a He-rich environment the mass-8 gap is bridged by either $\alpha +\alpha +\alpha \rightarrow^{12}$C
or $\alpha +\alpha + n \rightarrow ^9$Be reactions.
During this process, a neutron-rich freeze out occurs which triggers the r-process
\cite{Woo94}. At this stage, it would also be possible to bridge the $A = 8$ gap through the
reaction chain $^4$He(2n,$\gamma$)$^6$He(2n,$\gamma$)$^8$He($\beta^-$)$^8$Li(n,$\gamma$)$^9$Li($\beta^-$)$^9$Be
\cite{Efr96,Ros01}. This chain
provides an alternative path to proceed along the neutron-rich side of the line of stability
towards heavier isotopes.  One needs to know to what extent this
chain competes with the $^8$Li($\beta^-$)$^8$Be(2$\alpha$) process. An important clue to the answer depends
on an accurate knowledge of  the  $^8$Li(n,$\gamma$)$^9$Li reaction rate.

We consider the E1 $s$- and $d$-wave captures to both the ground and
the 1st excited state of $^{9}\text{Li}$. The $J_{b}=3/2^{-}$ ground
state and $J_{b}=1/2^{-}$ 1st excited state in $^{9}\text{Li}$ are
described as a $j_{b}=p_{3/2}$ neutron coupled to the
$^{8}\text{Li}$ core, which has an intrinsic spin $I_{x}=2^{+}$.
Here we use $a=0.52$ fm, $R=2.499$ fm and $V_{so}=-9.9$ MeV, which
are adopted from Ref. \cite{Bertulani99}. The spectroscopic factors
used in Ref. \cite{Mao90} are $1.65$ and $0.55$ for the ground and
1st excited state, respectively. However, for the ground state, most
of experiments and calculations give $SF \approx 0.8$ (see the
summary in Ref. \cite{Kanungo08}). Thus we use $SF=0.8$ instead of
$1.65$ for the ground state. The result is shown in Fig. \ref{8Lin}.
The experimental data are from Ref. \cite{ZGG98} using the Coulomb
dissociation of $^{9}\text{Li}$ on $\text{Pb}$ targets at 28.5 MeV/A
beam energy. From the result one can see the capture to the excited
state is much weaker than that to the ground state. Our ANC
($\sqrt{(SF)b^2}$) is $1.12$ fm$^{-1/2}$ for the ground state of
$^{9}$\textrm{Li} and $0.40$ fm$^{-1/2}$ for the 1st excited state
of $^{9}$\textrm{Li}.

\subsection{$^{11}$\textrm{B}$(\text{n},\gamma)^{12}\text{B}$}

Nucleosynthesis in inhomogeneous
big bang models are considerably dependent on neutron
capture reactions on light nuclei. Such reactions are also of
crucial relevance for the s-process nucleosynthesis in red
giant stars. To determine the reaction rates for
such different temperature conditions, the neutron capture cross sections need
to be known for a wide energy range.

Primordial nucleosynthesis might be affected by spatial variations
of both baryon-to-photon and neutron-to-proton ratios, the later being
caused by the short diffusion time for neutrons in the primordial plasma.
A possible signatures of baryon-number-inhomogeneous  big bang is the presence of
a  high primordial lithium abundance, or a high abundance of beryllium and boron isotopes.
As previously mentioned, inhomogeneous big bang models involve chain reactions such as
\cite{App88} $^1$H(n,$\gamma$)$^2$H(n,$\gamma$)$^3$H(d,n)$^4$He(t, $\gamma$)$^7$Li(n, $\gamma$)$^8$Li,
and  $^8$Li($\alpha$,n)$^{11}$B (n,$\gamma$)$^{12}$B($\beta^-$)$^{12}$C(n,$\gamma$)$^{13}$C, etc.,
paving the way to heavier nuclei. Thus, the reaction $^{11}$\textrm{B}$(\text{n},\gamma)^{12}\text{B}$ is
an important piece of inhomogeneous big bang scenarios \cite{Kaj90}.

The E1 $s$- and $d$-wave captures to the ground state of
$^{12}\text{B}$ are calculated. The $J_{b}=1^{+}$ ground state of
$^{12}\text{B}$ is described as a $j_{b}=p_{3/2}$ neutron coupled to
the $^{11}$\textrm{B} core, which has an intrinsic spin
$I_{x}=3/2^{-}$. Ref. \cite{Tsang05} extracts the ground state neutron spectroscopic factors for several light
by analyzing the previously reported measurements of the angular distributions in (d,p) and (p,d) reactions.
We adopt the spectroscopic factor $SF=1.09$ as in Ref.
\cite{Tsang05}. Our result for the non-resonant capture (solid line) is shown in Fig. \ref{11Bn}. The
experimental data are from Ref. \cite{IJV62}.

Similar to   Ref.
\cite{IJV62}, we describe the total capture cross section by a sum
of non-interfering Breit Wigner resonances superimposed on a slowly
varying background (non-resonant capture, solid line in the figure)
and the radiation widths of the levels are found to be 0.3 eV at
0.36 MeV, 0.3 eV at 0.87 MeV, 0.2 eV at 1.08 MeV, and 0.9 eV at 1.50
MeV, with estimated uncertainties of about 50\%.

Without comparison
to any experimental data, Ref. \cite{Lin03} describes a
calculation using a potential model, where captures to the second
and third excited states are considered. Their result is twice as
large as the experimental data of Ref. \cite{IJV62}.

In Ref. \cite{Liu01} the transfer reactions $^{11}$B(d,p)$^{12}$B and $^{12}$C(d,p)$^{13}$C,
at incident energy of 11.8 MeV, have been used to
extract the ANC for $^{12}$B $\rightarrow$ n + $^{11}$B. The ANC found
in Ref. \cite{Liu01}\ is $1.08$ fm$^{-1/2}$ ($C^{2}=1.16\pm0.10$
fm$^{-1}$). Our calculated ANC is $\sqrt{(SF)b^2}=1.41$ fm$^{-1/2}$.

\subsection{$^{12}$\textrm{C}($\text{n}$,
$\gamma$)$^{13}$\textrm{C}}

As mentioned above, not only the $^{11}$\textrm{B}$(\text{n},\gamma)^{12}\text{B}$,
but also the $^{12}$\textrm{C}($\text{n}$,
$\gamma$)$^{13}$\textrm{C} radiative capture is an important reaction in stellar
nucleosynthesis \cite{App88}.

We calculated the direct capture to the ground state and the first 3
excited states of $^{13}$C and compared with the experimental
results of Refs. \cite{Ohsaki94,Kikuchi95}. The $J_{b}=1/2^{-}$
ground state of $^{13}$\textrm{C} ($J_{b}=1/2^{+}$ for the 1st
excited state, $J_{b}=3/2^{-}$ for the 2nd excited state and
$J_{b}=5/2^{+}$ for the 3rd excited state) is described as a
$j_{b}=p_{1/2}$ neutron ($j_{b}=s_{1/2}$ neutron for the 1st excited
state, $j_{b}=p_{3/2}$ neutron for the 2nd excited state,
$j_{b}=d_{5/2}$ neutron for the 3rd excited state, respectively)
coupled to the $^{12}$\textrm{C} core, which has an intrinsic spin
$I_{x}=0^{+}$. In this particular case, we use $r_{0}=1.236$ fm,
$a=0.62$ fm and $V_{so}=-7$ MeV. These are the same set of
parameters adopted in Ref. \cite{Mengoni95}. The spectroscopic
factors published in Ref. \cite{Ajzenberg91} are  $SF=0.77$ for the
ground state, $SF=0.65$ for the 1st excited state, $SF=0.14$ for the
2nd excited state, and $SF=0.58$ for the 3rd excited state. We adopt
these values, except for the 1st excited state. For this state, we
use $SF=0.8$ because it yields a better description of the
experimental data in our model. It is also the same value adopted in
Ref. \cite{Kikuchi95}.

It is also necessary to vary the potential depth for the
continuum states for transitions to the different bound states in $^{13}$C.
For the capture to the 1st and 3rd excited states, we use $V_c=V_b$, where $V_b$ are
used to describe the neutron separation energies of the two excited states in $^{13}$C (see Table III).
For the capture to the ground state we use
$V_{c}=-14.75$ MeV, whereas for the capture to the 2nd excited state, $V_{c}=-11.50$ MeV is adopted. Our
results are shown in Fig. \ref{12Cn}.
This reaction has
also been studied in Refs. \cite{Mengoni95, Kikuchi95, Baye04,
Lin03} where a variety of potential models have been used and different
spectroscopic factors were adopted.

Our calculated ANC is
$\sqrt{(SF)b^2}=1.62$ fm$^{-1/2}$ for the ground state and
$1.61$ fm$^{-1/2}$, $0.23$ fm$^{-1/2}$, and $0.11$ fm$^{-1/2}$ for
the 1st, 2nd and 3rd excited states, respectively.
In Ref. \cite{Liu01} the transfer reactions $^{11}$B(d,p)$^{12}$B and $^{12}$C(d,p)$^{13}$C,
at incident energy of 11.8 MeV, have been used to
extract the ANC for $^{13}$C $\rightarrow$ n + $^{12}$C.
The ANC found in
Ref. \cite{Lin03} for the 1st excited is $1.84\pm0.16$ fm$^{-1/2}$, in close
agreement with our 1.61 fm$^{-1/2}$ value.

\subsection{$^{14}\text{C}$($\text{n}$,$\gamma$)$^{15}$\textrm{C}}
As we have discussed previously, inhomogeneous big bang models allow for
the synthesis of heavy elements via a chain of neutron capture reactions. This
includes the $^{14}\text{C}$($\text{n}$,$\gamma$)$^{15}$\textrm{C} reaction.
Nucleosynthesis depends on reactions that destroy
$^{14}$C, the most important of which is $^{14}\text{C}$($\text{n}$,$\gamma$)$^{15}$\textrm{C}.
This reaction is also a part of the neutron induced CNO cycles in
the helium burning layer of AGB stars,
in the helium burning core of massive stars, and in subsequent
carbon burning \cite{Wie99}. Such cycles may cause a
depletion in the CNO abundances. The $^{14}\text{C}$($\text{n}$,$\gamma$)$^{15}$\textrm{C} reaction
is the slowest of both of these cycles and, therefore the
knowledge of its rate is important to predict the $^{14}$C abundances.

Due to the weak binding of the $^{15}$C ground state, and because
there are no low lying resonances, the cross section is mainly
determined by an $E1$ non-resonant transition from an initial p-wave
scattering state to the ground state \cite{Timofeyuk06}. The
$J_{b}=1/2^{+}$ ground state of $^{15}\text{C}$ is described as a
$j_{b}=s_{1/2}$ neutron coupled to the $^{14}$\textrm{C} core, which
has an intrinsic spin $I_{x}=0^{+}$.

In Ref. \cite{Goss75} a 14 MeV deuteron beam was used to measure the angular distributions for the $^{14}$C(d, p)$^{15}$C
reaction leading to the two bound states and eight of the unbound states of $^{15}$C. An spectroscopic factor
$SF=0.88$ for the ground state of $^{14}$C has been inferred. Adopting this value, we obtain the
DC cross section shown in Fig. \ref{14Cn}. The experimental data are from
Ref. \cite{Reifarth08}.

In Ref. \cite{Summers08} a theoretical analysis of existing experimental data on the Coulomb dissociation of $^{15}$C on 208Pb at 68 MeV/nucleon was used to infer the asymptotic normalization coefficient for $^{15}$C $\rightarrow$ n + $^{14}$C. The ANC value reported in Ref. \cite{Summers08}\ is $1.13$ fm$^{-1/2}$ ($C^{2}%
=1.28\pm0.01$ fm$^{-1}$). Our ANC value is $\sqrt{(SF)b^2}=1.35$
fm$^{-1/2}$.

\subsection{$^{15}$\textrm{N}$(n,\gamma)^{16}$\textrm{O}}
The cross section for the reaction $^{15}$N(n,$\gamma$)$^{16}$N is an important input in
the reaction network for the production of heavier isotopes in both inhomogeneous big bang and in
red giant environments \cite{App88}.

The direct capture for this reaction is dominated by
the $p\rightarrow d$ wave transition to the ground state,
$p\rightarrow s$ wave transition to the first excited state of $^{16}$N at
$0.120$ MeV, $p\rightarrow d$ wave transitions to the second excited
state at $0.296$ MeV and $p\rightarrow s$ wave transitions to the
third excited state at $0.397$ MeV. These conclusions were made in Ref.  \cite{MSH96},
where reaction cross
sections of $^{15}$\textrm{N}$(n,\gamma)^{16}$\textrm{O} was reported and direct capture and shell model
calculations were performed to interpret their data.  The gamma-ray transitions are
all dominated by the $E1$ multipolarity. The $J_{b}=2^{-}$ ground
state ($J_{b}=0^{-}$ 1st excited state, $J_{b}=3^{-}$ 2nd excited
state, $J_{b}=1^{-}$ 3rd excited state) $^{16}\text{N}$ is described
as a $j_{b}=d_{5/2}$ neutron ($j_{b}=s_{1/2}$ neutron,
$j_{b}=d_{5/2}$ neutron, $j_{b}=s_{1/2}$ neutron) coupled to the
$^{15}$\textrm{N} core, which has an intrinsic spin $I_{x}=1/2^{-}$.

In Ref.
\cite{Bohne72}  (d,n) and (d,p) reactions on $^{15}$N were measured
and Hauser-Feshbach calculations
were used to  extract spectroscopic factors with $30\%$ uncertainty. Their values are
$SF=0.55$ for the ground
state, $SF=0.46$ for the $2^{-}$ state, $SF=0.54$ for the $3^{-}$
state and $SF=0.52$ for the $1^{-}$ state. Our result is shown in Fig. \ref{15Nn}.
The experimental data are
from Ref. \cite{MSH96}.  Our calculations yield
similar results as those of Ref. \cite{MSH96} and
Ref. \cite{Herndl99}, and reproduce the experimental data
rather well, considering the $\pm 30\%$ error in the spectroscopic factor (see dashed
line in Fig. \ref{15Nn}).

Our calculated ANCs  are
$0.85$ fm$^{-1/2}$ for the ground state of $^{9}$\textrm{Li}, $1.10$
fm$^{-1/2}$ for the first excited state, $0.29$ fm$^{-1/2}$ for the
second excited state and $1.08$ fm$^{-1/2}$ for the third excited
state, respectively.

\subsection{$^{16}$\textrm{O}$(n,\gamma)^{17}\text{O}$}

This reaction is important for s-processes for various metallicity stars and for inhomogeneous big
bang models, which, for masses beyond $A>12$ can proceed via $^{12}$C(n,$\gamma$)$^{13}$C(n,$\gamma$)$^{14}
$C(n,$\gamma$)$^{15}$N(n,$\gamma$)$^{16}$N($\beta^-$)$^{16}$O(n,$\gamma$)$\dots$

The non-resonant, direct capture, to the ground
state and to the 1st excited state of $^{17}\text{O}$ dominates
the cross section in the energy range of $0.02-0.28$ MeV  \cite{INM95}. The
gamma-ray transitions are dominated by the $E1$ multipolarity and by
incoming $p$-waves. The $J_{b}=5/2^{+}$ ground state
($J_{b}=1/2^{+}$ 1st excited state) of $^{17}\text{O}$ is
described as
a $j_{b}=d_{5/2}$ neutron ($j_{b}=s_{1/2}$ neutron) coupled to the $^{16}%
$\textrm{O} core, which has an intrinsic spin $I_{x}=0^{+}$. We use
a spectroscopic factor $SF=1.0$ for both ground and excited states.

The results of our calculations for these two captures are shown in the top panel of
Fig. \ref{16On} separately. The experimental data are from Ref.
\cite{INM95}. Our potential model calculations yield similar results as the
calculations Ref.
\cite{Dufour01}, where a microscopic multicluster model was used.
The total cross section is shown in the bottom panel of Fig.
\ref{16On} together with a theoretical result from Ref. \cite{CKH08} where
direct and semi-direct
components of the neutron capture cross sections were calculated.

Our
calculated ANC ($\sqrt{(SF)b^2}$) is $0.90$ fm$^{-1/2}$ for the
ground state of $^{17}\text{O}$ and $3.01$ fm$^{-1/2}$ for the 1st
excited state of $^{17}\text{O}$.

\subsection{$^{18}$\textrm{O}$(\text{n},\gamma)^{19}\text{O}$}

Further nucleosynthesis during inhomogeneous big bang models towards higher
masses is controlled by the reaction rate of $^{18}$\textrm{O}$(\text{n},\gamma)^{19}\text{O}$.
If this reaction is stronger than the $^{18}$O(p,$\alpha$)$^{15}$N reaction
then material be processed out of the CNO cycle to the
region above $A>20$. This reaction is also of interest for
stellar helium burning in AGB stars by means of s-processes.

The direct capture for this
reaction is dominated by $p\rightarrow d$-wave transitions to the
ground state, the first excited state at $0.096$ MeV, and the
$p\rightarrow s$ transition to the second excited state at $1.47$
MeV  \cite{Meissner96}. The gamma-ray transitions are all dominated by the $E1$
multipolarity. The $J_{b}=5/2^{+}$ ground state ($J_{b}=3/2^{+}$ 1st
excited state, $J_{b}=1/2^{+}$ 2nd excited state) of $^{17}\text{O}$
is described as a $j_{b}=d_{5/2}$ neutron ($j_{b}=d_{3/2}$ neutron,
$j_{b}=s_{1/2}$ neutron) coupled to the $^{18}$\textrm{O} core,
which has an intrinsic spin $I_{x}=0^{+}$.

We have adopted spectroscopic factors
from Ref. \cite{Meissner96}. They are $SF=0.69$ for the
ground state, $SF=0.013$ for the $3/2+$ state, and $SF=0.83$ for the
$1/2^{+}$ state. Our results are shown in Fig. \ref{18On}. They are
close to the calculations reported in  Refs. \cite{Meissner96,Herndl99}. The experimental data are from Ref.
\cite{Meissner96}. The data points at $0.138$ MeV and $0.331$ MeV
are much higher than our non-resonant calculation because of the
resonances at $0.152$ MeV and $0.371$ MeV, corresponding to the
$3/2^{+}$ state at $4.109$ MeV and to the state at $4.328\pm003$ MeV
in $^{19}$\textrm{O}, respectively. This has been discussed in details in Ref.
\cite{Meissner96}.

Our calculated ANC ($\sqrt{(SF)b^2}$) is $0.75$
fm$^{-1/2}$ for the ground state of $^{19}\text{O}$, $0.09$
fm$^{-1/2}$ for the first excited state and $2.26$ fm$^{-1/2}$ for
the second excited state.

\section{Sensitivity on the potential depth parameter}

As with any other  model, the results obtained with the
single-particle model for the cross sections can be very sensitive to the
choice of  parameters. In order to check this sensitivity, in
Table \ref{t4} we compare the cross sections at $0.4$ MeV for the
capture to the ground state of the reaction
$^{16}$\textrm{O}(\textrm{p}, $\gamma$)$^{17}$\textrm{F} with that
of $^{16}$\textrm{O}(\textrm{n}, $\gamma$)$^{17}$\textrm{O}. The
potential depth for continuum state $V_{c}$ has been varied by
$\pm10\%$ to test the sensitivity of the cross sections on $V_{c}$.%

The $V_{c}$ in the third (last) column is $10\%$ smaller (larger)
than that of the fourth column, which is used in the calculation for
the S-factors or cross sections in sections III and IV. From  table
\ref{t4}, one can conclude that proton capture is less sensitive to
the internal part of the potential, as expected.  This is due to the
Coulomb barrier. In other words, proton capture reactions tend to be
more peripheral than neutron capture reactions. In the proton
capture case, the ANC technique is thus expected to work better than
in the neutron capture one. But these conclusions obviously change
in the presence of potential resonances, when the cross sections can
suddenly change by orders of magnitude if the potential depth is
slightly varied.

In order to show the large sensitivity of the S-factor, or cross
section, on potential parameters close to a resonance, we use the
test-case of the $^{15}$N(p,$\gamma$) reaction. This is shown in
figure \ref{s15N} where we plot the ratio between the S-factor at
$E=0$ calculated with a potential depth $V_c$ and the S-factor
calculated with a zero potential depth: $S(0,V_c)/S(0,0)$. The open
circle corresponds to the value of $V_{c}$ used in the calculation
presented in figure \ref{15Np}.

As is clearly seen in figure \ref{15Np}, a small change
(i.e. by 10\%) in the value of $V_c$ can cause orders of magnitude
change in the corresponding S-factor near a resonance. Thus,
although one can indeed reproduce resonant states with the potential
model, one has to be very careful with the values of observables
obtained with the model, such as the ANCs, or spectroscopic factors.
These will also be over-sensitive to the potential fitting
parameters.

\section{ANCs from single-particle models}
In figure \ref{Ratio} we show the ratio of our calculations of ANCs
($\sqrt{(SF) b^{2}}$) with the ANCs extracted from the literature  and mentioned in
this article.  Not all ANCs are shown because either they have not been
indirectly extracted from experiments, or calculated previously. The solid circles are for proton capture whereas the
solid triangles are for neutron capture. The dashed line is a guide to the eye and shows the
ratio equal to unity. We notice that our ANCs differ up to a factor of
1.6 from previously reported values.

In our calculations, the ANCs  are indirectly obtained
by adjusting our calculated S-factors or cross sections to the
available experimental data. The ANC's from literature are partially obtained
by indirectly fitting calculations to experimental data  in transfer reactions,
or by means of elaborate microscopic models, or else.
Evidently, a more consistent comparison between these values
deserves a more detailed study.

\medskip

\section{Final conclusions}

In this article, we have explored the single-particle potential model to
describe radiative proton and neutron capture reactions of
relevance for astrophysics. Using a well defined approach
and the same numerical code, we have obtained spectroscopic factors
and single-particle asymptotic normalization coefficients for
several reactions in the mass range $A<20$.

We have only considered cases for which potential models
yields reasonable results. There are several radiative capture
reactions which do not fall into this category. They require a more derailed
study, with possible adjustments and/or extensions of the
model. Evidently, there will be situations
for which the potential model will always fail.

Our work
has shown minor differences with previously published
results. We have demonstrated that there is a reasonable
justification for the use of  potential model calculations for many
reactions which have either been measured experimentally, or
calculated theoretically.

A
systematic study of asymptotic normalization coefficients and
spectroscopic factors based on the single-particle model is very
useful to validate other theoretical descriptions of radiative
capture reactions. This study is also relevant to
correlate spectroscopic observables to other nuclear properties.
Work in this direction is also in progress.

\medskip This work was partially supported by the U.S. DOE grants
DE-FG02-08ER41533 and DE-FC02-07ER41457 (UNEDF, SciDAC-2).

\newpage
\noindent\uppercase{\textbf{Tables and Graphs}}
\begin{table*}[ph]
\begin{center}
\begin{tabular}
[c]{|c|c|}\hline Parameter & Adopted value\\\hline
$R_{0}=R_{S0}=R_{C}$ & $r_{0}(A+1)^{1/3}$ fm\\\hline $r_{0}$ &
$1.25$\\\hline $a_{0}=a_{S0}$ & $0.65$ fm\\\hline $V_{s0}$ & $-10$
MeV\\\hline
\end{tabular}
\end{center}
\caption{Parameters of the single-particle potentials, except for
few cases explicitly mentioned in the text.}%
\label{WS}
\end{table*}

\begin{table*}[hp]
\begin{center}
\begin{tabular}{|c|c|c|c|c|c|c|}
\hline Reaction & $E_{b}$ & $V_{b}$ & $SF$ & $b$ & $>R_{0}$ & $S(0)$
\\ \hline \textrm{d}$(\text{p},\gamma )^{3}\text{He}$ & 5.49 &
-44.43 & 0.7 & 1.86 & 0.98 & $0.14$ \\ \hline
$^{6}$\textrm{Li}$($\textrm{p}$,\gamma )^{7}\text{Be}$ & 5.61 &
-65.91 & \multicolumn{1}{|l|}{0.83 \cite{Camargo08}} & 2.21 & 1.28 &
66.8 \\ \hline $^{6}$\textrm{Li}$($\textrm{p}$,\gamma
)^{7}\text{Be}^{\ast }$ & 5.18 &
-64.94 & \multicolumn{1}{|l|}{0.84 \cite{Arai02}} & 2.08 & 1.19 & 32.7 \\
\hline $^{7}$\textrm{Li}(\textrm{p}, $\gamma $)$^{8}$\textrm{Be} &
17.26 & -75.69 & 1.0 & 7.84 & 1.01 & 238. \\ \hline
$^{7}\text{Be}(\text{p},\gamma )^{8}\text{B}$ & 0.14 & -41.26 & 1.0
& 0.72 & 1.00 & 19.4 \\ \hline $^{8}$\textrm{B}(\textrm{p}, $\gamma
$)$^{9}$\textrm{C} & 1.30 & -41.97 & 1.0 \cite{Mohr03} & 1.31 & 1.08
& 42.5 \\ \hline
$^{9}\text{Be}(\text{p},\gamma )^{10}\text{B}$ & 6.59 & -49.83 & 1.0 \cite%
{Miura92} & 3.43 & 1.27 & 1052 \\ \hline
$^{11}$\textrm{C}(\textrm{p}, $\gamma $)$^{12}$\textrm{N} & 0.60 &
-40.72 & 0.4 \cite{Lefebvre95} & 1.49 & 1.01 & 50.8 \\ \hline
$^{12}$\textrm{C}(\textrm{p}, $\gamma $)$^{13}$\textrm{N} & 1.94 &
-41.65 & 1.0 & 2.05 & 1.04 & 2346 \\ \hline
$^{13}$\textrm{C}(\textrm{p}, $\gamma $)$^{14}$\textrm{N} & 7.55 &
-50.26 & 0.33 & 5.31 & 1.10 & 6217 \\ \hline
$^{13}$\textrm{N}(\textrm{p}, $\gamma $)$^{14}$\textrm{O} & 4.63 &
-46.02 & 1.88 \cite{Li06} & 3.97 & 1.45 & 5771 \\ \hline
$^{14}\text{N}(\text{p},\gamma )^{15}$\textrm{O}$^{\ast }$ & 0.50 &
-14.83 & 1.5 & 4.24 & 1.00 & 1470 \\ \hline
$^{15}$\textrm{N}$(\text{p},\gamma )^{16}\text{O}$ & 12.13 & -54.81
& 1.8 \cite{RR74} & 10.16 & 0.78 & $2.21 \cdot 10^4$ \\ \hline
$^{16}\text{O}(\text{p},\gamma )^{17}\text{F}$ & 0.60 & -49.69 & 0.9 \cite%
{Rolfs73} & 0.96 & 1.02 & 304 \\ \hline
$^{16}\text{O}(\text{p},\gamma )^{17}\text{F}^{\ast }$ & 0.11 &
-50.70 & 1.0 \cite{Rolfs73} & 77.21 & 1.00 & 9075 \\ \hline
$^{20}\text{Ne}(\text{p},\gamma )^{21}\text{Na}^{\ast }$ & 0.006 &
-47.24 & 0.7 & 4.02 & 1.00 & $4.28\cdot 10^4$ \\ \hline
$^{20}\text{Ne}(\text{p},\gamma )^{21}\text{Na}^{\ast }$ & 2.10 &
-49.63 & 0.8 & 2.43 & 1.00 & 2493 \\ \hline
\end{tabular}
\end{center}
\caption{Binding energy ($E_{b}$, in \textrm{MeV}), central
potential depth of bound state ($V_{b}$, in \textrm{MeV}),
spectroscopic factor ($SF$), single-particle asymptotic
normalization coefficients ($b$, in fm$^{-1/2}$), the factor that
multiplies S-factor if the integration in     Eq. \protect
\ref{respf} starts at $r=R_{0}$ (nuclear radius) and S-factor at
zero energy ($S(0)$, in eV b) for radiative proton capture
reactions.} \label{t2}
\end{table*}

\begin{table*}[tbp]
\begin{center}
\begin{tabular}{|c|c|c|c|c|c|}
\hline Reaction & $E_{b}$ & $V_{b}$ & $SF$ & $b$ & $r>R_{0}$ \\
\hline $^{2}$\textrm{H(}$\text{n},\gamma
$\textrm{)}$^{3}$\textrm{H} & 6.26 & -44.63 & 1.0 & 1.90 & 0.97 \\
\hline $^{7}\text{Li}$\textrm{(}$\text{n},\gamma
$\textrm{)}$^{8}\text{Li}$ & 2.03 & -43.56 & 0.87 \cite{Nagai} &
0.76 & 1.04 \\ \hline
$^{7}\text{Li}$\textrm{(}$\text{n},\gamma $\textrm{)}$^{8}\text{Li}%
^{\ast }$ & 1.05 & -40.46 & 0.48 \cite{Nagai} & 0.47 & 1.02 \\
\hline $^{8}\text{Li(n},\gamma \text{)}^{9}\text{Li}$ & 4.06 &
-45.29 & 0.8 \cite{Kanungo08} & 1.25 & 1.08 \\ \hline
$^{8}\text{Li(n},\gamma \text{)}^{9}\text{Li}^{\ast }$ & 1.37 &
-38.57 & 0.55 \cite{Mao90} & 0.54 & 1.03 \\ \hline
$^{11}$\textrm{B(}$\text{n},\gamma $\textrm{)}$^{12}\text{B}$ & 3.37
& -34.33 & 1.09 \cite{Tsang05} & 1.35 & 1.09 \\ \hline
$^{12}$\textrm{C}(\textrm{n}, $\gamma $)$^{13}$\textrm{C} & 4.95 &
-41.35 & 0.77 \cite{Ajzenberg91} & 1.85 & 3.23 \\ \hline
$^{12}$\textrm{C}(\textrm{n}, $\gamma $)$^{13}$\textrm{C}$^{\ast }$
& 1.86 & -56.90 & 0.8 \cite{Kikuchi95} & 1.80 & 1.00 \\ \hline
$^{12}$\textrm{C}(\textrm{n}, $\gamma $)$^{13}$\textrm{C}$^{\ast }$
& 1.27 & -28.81 & 0.14 \cite{Ajzenberg91} & 0.61 & 1.23 \\ \hline
$^{12}$\textrm{C}(\textrm{n}, $\gamma $)$^{13}$\textrm{C}$^{\ast }$
& 1.09 & -56.85 & 0.58 \cite{Ajzenberg91} & 0.15 & 1.04 \\ \hline
$^{14}\text{C(n},\gamma \text{)}^{15}$\textrm{C} & 1.22 & -48.63 &
0.88 \cite{Goss75} & 1.44 & 1.00 \\ \hline $^{15}\text{N(n},\gamma
\text{)}^{16}$\textrm{N} & 2.49 & -27.06 & 0.55 \cite{Bohne72} &
1.14 & 1.38 \\ \hline $^{15}\text{N(n},\gamma
\text{)}^{16}$\textrm{N}$^{\ast }$ & 2.37 & -12.45 & 0.46
\cite{Bohne72} & 1.62 & 1.11 \\ \hline $^{15}\text{N(n},\gamma
\text{)}^{16}$\textrm{N}$^{\ast }$ & 2.19 & -49.51 & 0.54
\cite{Bohne72} & 0.39 & 2.77 \\ \hline $^{15}\text{N(n},\gamma
\text{)}^{16}$\textrm{N}$^{\ast }$ & 2.09 & -11.90 & 0.52
\cite{Bohne72} & 1.50 & 0.94 \\ \hline
$^{16}$\textrm{O(}$\text{n},\gamma $\textrm{)}$^{17}\text{O}$ & 4.14
& -51.77 & 1.0 & 0.90 & 1.17 \\ \hline
$^{16}$\textrm{O(}$\text{n},\gamma $\textrm{)}$^{17}\text{O}^{\ast
}$ & 3.27 & 51.60 & 1.0 & 3.01 & 0.99 \\ \hline
$^{18}$\textrm{O(}$\text{n},\gamma $\textrm{)}$^{19}\text{O}$ & 3.96
& -47.79 & 0.69 \cite{Meissner96} & 0.90 & 1.17 \\ \hline
$^{18}$\textrm{O(}$\text{n},\gamma $\textrm{)}$^{19}\text{O}^{\ast
}$ & 3.86 & -55.94 & 0.013 \cite{Meissner96} & 0.81 & 1.14 \\ \hline
$^{18}$\textrm{O(}$\text{n},\gamma $\textrm{)}$^{19}\text{O}^{\ast
}$ & 2.49 & -46.33 & 0.83 \cite{Meissner96} & 2.48 & 1.00 \\ \hline
\end{tabular}
\end{center}
\caption{Binding energy ($E_{b}$, in \textrm{MeV}), central
potential depth of bound state ($V_{b}$, in \textrm{MeV}),
spectroscopic factor ($SF$), single-particle asymptotic
normalization coefficients ($b$, in fm$^{-1/2}$) and the factor
multiplying the S-factor assuming that the integration in     Eq.
\protect \ref{respf} starts at $r=R_{0}$ (nuclear radius).}
\label{t3}
\end{table*}

\begin{table*}[ptb]
\begin{center}
\begin{tabular}
[c]{|c|c|c|c|c|}\hline
\multicolumn{1}{|c|}{} & $V_{c}$ (MeV) & $44.72$ & $49.69$ & $54.66$%
\\\cline{2-5}%
\multicolumn{1}{|c|}{$^{16}$\textrm{O}(\textrm{p},
$\gamma$)$^{17}$\textrm{F}} & $\sigma$ ($\mu b$) &
$4.63\times10^{-3}$ & $4.83\times10^{-3}$ &
$5.05\times10^{-3}$\\\cline{2-5}%
\multicolumn{1}{|c|}{} & $\Delta\sigma/\sigma$ & $-4.14\%$ &  & $+4.55\%$%
\\\hline
\multicolumn{1}{|c|}{} & $V_{c}$ (MeV) & $46.59$ & $51.77$ & $56.94$%
\\\cline{2-5}%
\multicolumn{1}{|c|}{$^{16}$\textrm{O}(\textrm{n},
$\gamma$)$^{17}$\textrm{O}}
& $\sigma$ ($\mu b$) & $14.35$ & $21.41$ & $38.42$\\\cline{2-5}%
\multicolumn{1}{|c|}{} & $\Delta\sigma/\sigma$ & $-32.98\%$ &  &
$+79.45\%$\\\hline
\end{tabular}
\caption{Cross sections at $0.4$ MeV for the capture to the ground
state of the reaction $^{16}$\textrm{O}(\textrm{p},
$\gamma$)$^{17}$\textrm{F} with that of
$^{16}$\textrm{O}(\textrm{n}, $\gamma$)$^{17}$\textrm{O}.}
\end{center}
\label{t4}
\end{table*}

\newpage
\begin{figure*}[ptb]
\begin{center}
\includegraphics[
height=2.4 in, width=3.37 in]{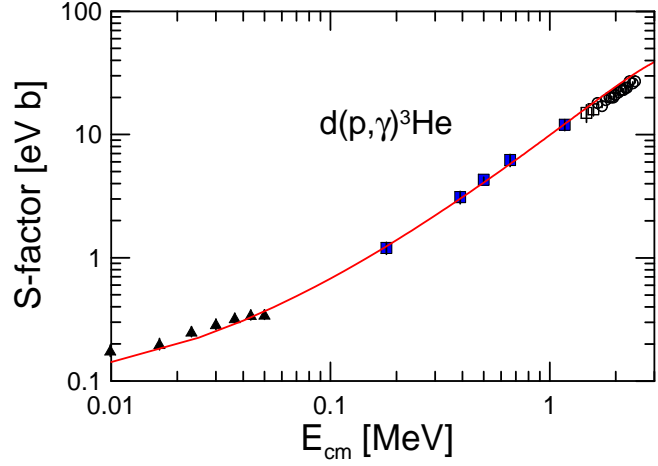}
\end{center}
\caption[$\textrm{d}(p,\gamma)^{3}\textrm{He}$]{(Color online).
Single-particle model calculation for the reaction $\textrm{d}
(p,\gamma)^{3}\textrm{He}$. Experimental data are from  Refs.
\cite{GLR62, BKS64, WBL67, SCL95}. The parameters calculated
according to Table I are used. The potential depth (here $V_b=V_c$)
is given in Table II.} \label{dp}
\end{figure*}

\begin{figure*}[ptb]
\begin{center}
\includegraphics[
height=2.4 in, width=3.45 in]{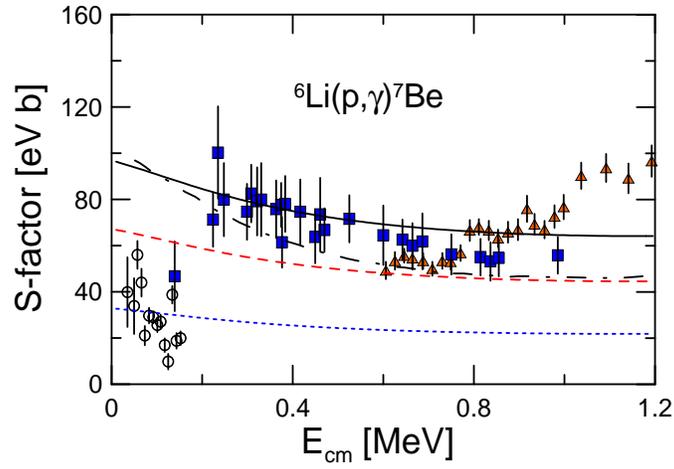}
\end{center}
\caption[$^{6}$\textrm{Li}$($\textrm{p}$,\gamma)^{7}\text{Be}$]{(Color online). Single-particle model calculation for the reaction $^{6}$\textrm{Li}%
$($\textrm{p}$,\gamma)^{7}\text{Be}$. The dotted line is the
calculation for the capture to the 1st excited of $^{7}\text{Be}$
and the dashed line for the ground state. The solid line is the
total calculated S-factor. Experimental data are from Refs.
\cite{Bruss93, Switkowski79, Arai02}. The dotted-dashed line is the
total S-factor calculated in
Ref. \cite{Arai02} using a four-cluster microscopic model.}%
\label{6Lip}%
\end{figure*}

\begin{figure*}[ptb]
\begin{center}
\includegraphics[
height=2.4 in, width=3.37 in  ]{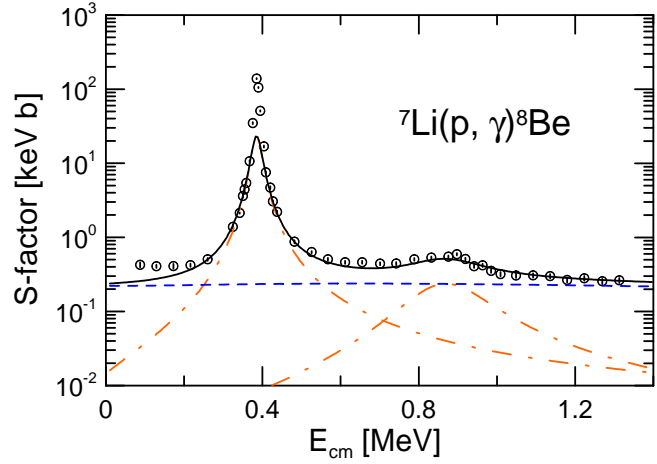}
\end{center}
\caption{(Color online). Potential model calculation for the
reaction $^{7}$\textrm{Li} (\textrm{p}, $\gamma$)$^{8}$\textrm{Be}.
Experimental data are from  Ref.
\cite{Zahnow95}. }%
\label{7Lip}%
\end{figure*}

\begin{figure*}[ptb]
\begin{center}
\includegraphics[
height=2.4 in, width=3.37 in ]{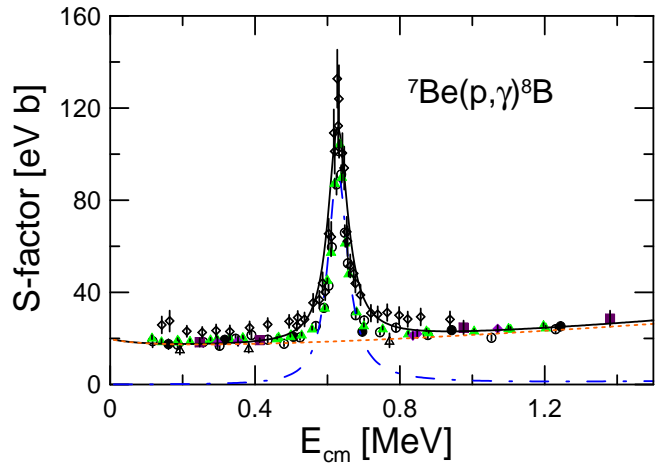}
\end{center}
\caption{(Color online). Single-particle model calculations for the reaction $^{7}$\textrm{Be}%
$(p,\gamma)^{8}\text{B}$. The dashed-dotted line is the calculation
for the M1 resonance at $E_{cm}=0.63$ MeV and the dotted line is for
the non-resonant capture. Experimental data are from Refs.
\cite{VCK70, FED83, Baby03, Junghans03, Iwasa99,Kav69}. The total S
factor is shown as a
solid line. }%
\label{7Bep}%
\end{figure*}

\begin{figure*}[ptb]
\begin{center}
\includegraphics[
height=2.4 in, width=3.37 in ]{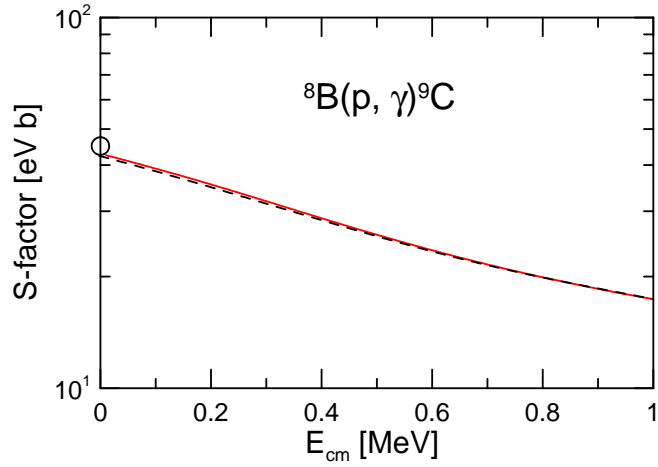}
\end{center}
\caption{(Color online). Single-particle model calculations for the reaction $^{8}$\textrm{B}%
(\textrm{p}, $\gamma$)$^{9}$\textrm{C }(solid line). The open circle
at $E=0$ is from  Refs.  \cite{Trache02, Beaumel01}. The result from
Ref.
\cite{Mohr03} ($\lambda_{scatt}=0.55$ fm) is shown as a dashed line.}%
\label{8Bp}%
\end{figure*}

\begin{figure*}[ptb]
\begin{center}
\includegraphics[
height=2.4 in, width=3.45 in ]{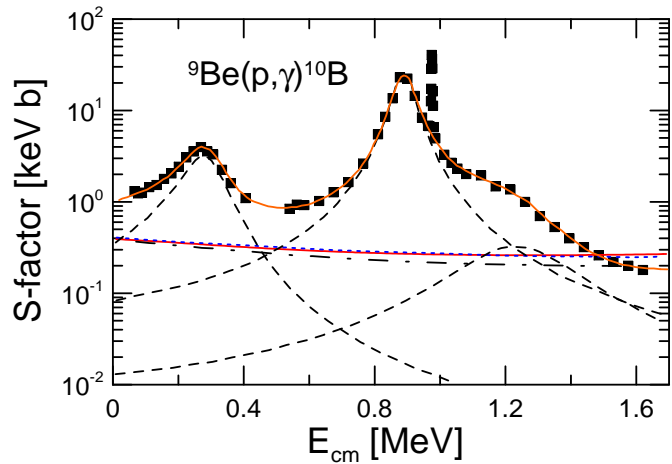}
\end{center}
\caption{(Color online). Single-particle model calculations for the reaction $^{9}$\textrm{Be}%
$(\text{p},\gamma)^{10}\text{B}$ (solid line). The experimental data
are from Ref. \cite{ZAJ95}. The fits to the resonances, done in Ref.
\cite{ZAJ95}, are shown as dashed lines. DC results from Ref.
\cite{Sattarov99} and Ref. \cite{ZAJ95} are shown as a dotted-dashed
line and a dotted line, respectively. The curve passing through the
experimental
data points is the sum of our DC calculation and the resonance fits, given by the dashed lines.}%
\label{11Bep}%
\end{figure*}

\begin{figure*}[ptb]
\begin{center}
\includegraphics[
height=2.4 in, width=3.37 in ]{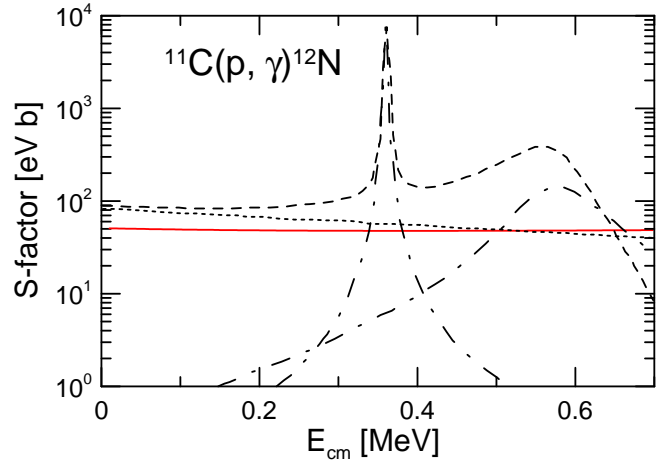}
\end{center}
\caption{(Color online). Single-particle  model calculations for the reaction $^{11}$\textrm{C}%
(\textrm{p}, $\gamma$)$^{12}$\textrm{N }(solid line). R-matrix
results from Ref. \cite{Tang03} are also shown by dashed lines
(resonances) and a
dotted line (non-resonant).}%
\label{11Cp}%
\end{figure*}

\begin{figure*}[ptb]
\begin{center}
\includegraphics[
height=2.4 in, width=3.48 in ]{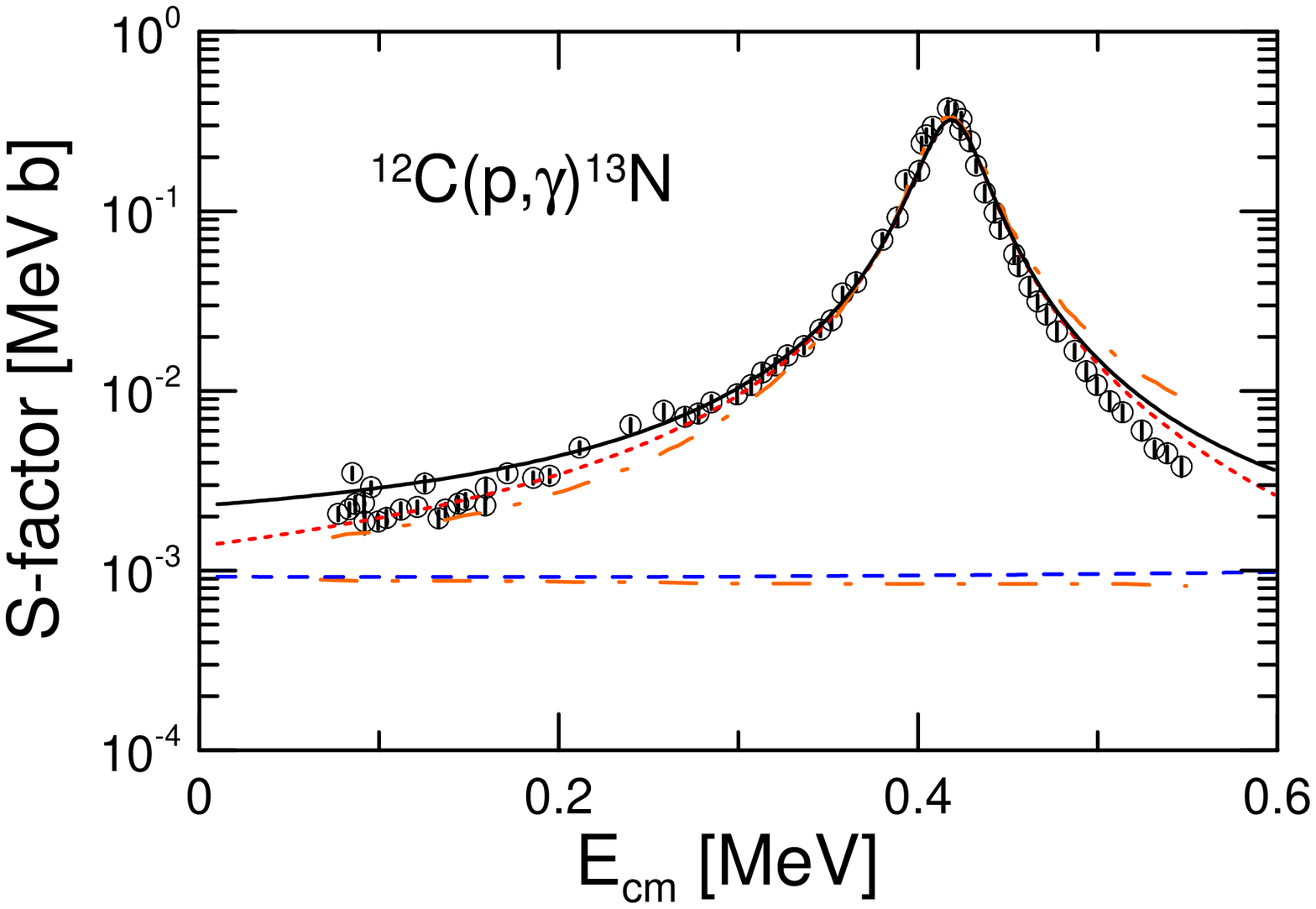}
\end{center}
\caption{(Color online). Single-particle  model calculations for the reaction $^{12}$\textrm{C}%
(\textrm{p}, $\gamma$)$^{13}$\textrm{N } are shown as a dashed line
(DC), a dotted line ($E1$ resonance) and a solid line (total). The
experimental data are from Ref. \cite{Rolfs74}. The potential model
results
from Ref. \cite{Nesaraja01} are shown as dotted-dashed lines.}%
\label{12Cp}%
\end{figure*}

\begin{figure*}[ptb]
\begin{center}
\includegraphics[
height=2.4 in, width=3.37 in ]{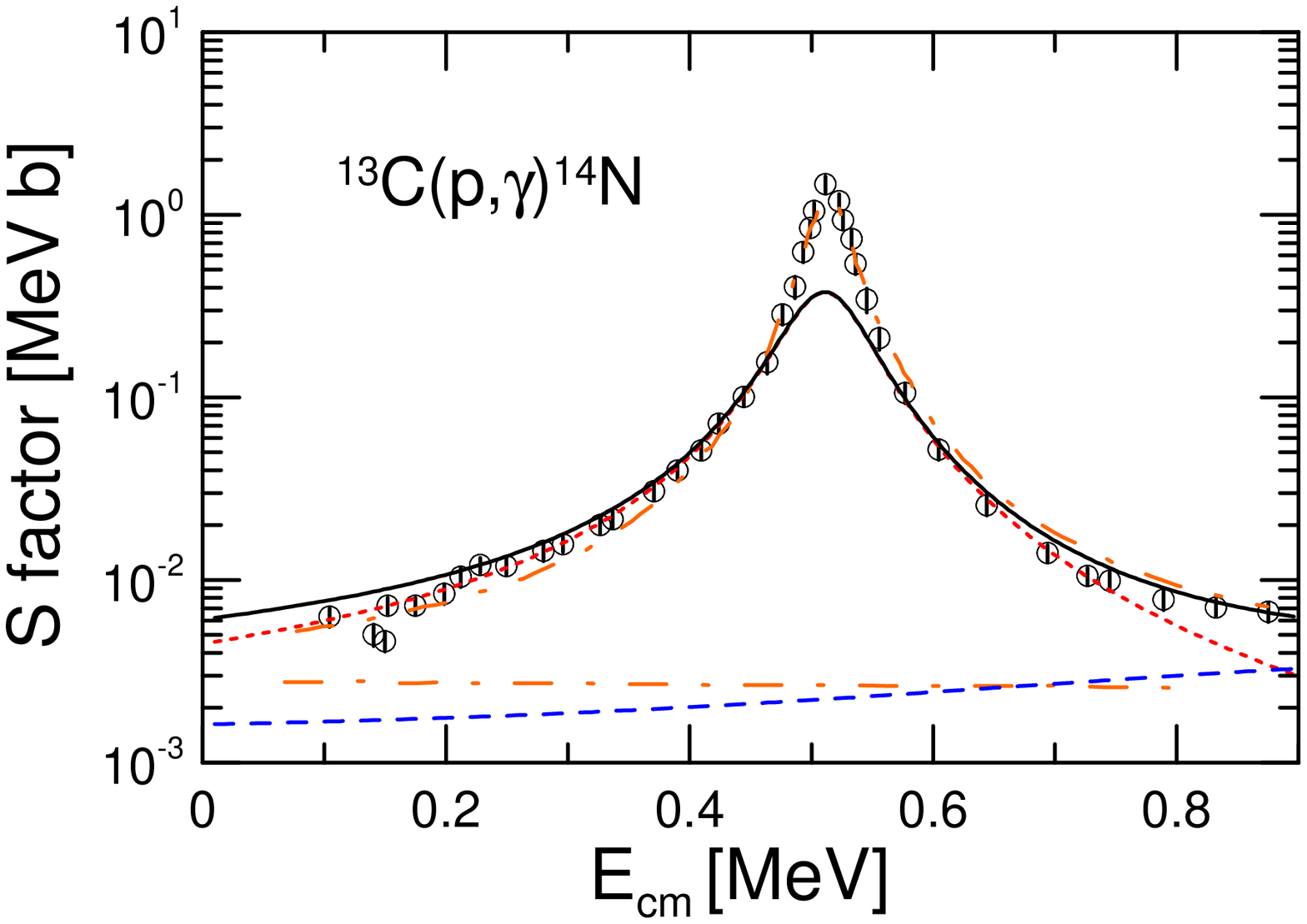}
\end{center}
\caption{(Color online). Single-particle  model calculations for the reaction $^{13}$\textrm{C}%
(\textrm{p}, $\gamma$)$^{14}$\textrm{N} are shown as a dashed line
(DC), a dotted line ($E1$ resonance) and a solid line (total). The
experimental data are from Ref. \cite{King94}. The potential model
results
from Ref. \cite{Nesaraja01} are shown as dotted-dashed lines.}%
\label{13Cp}%
\end{figure*}

\begin{figure*}[ptb]
\begin{center}
\includegraphics[
height=2.4 in, width=3.37 in ]{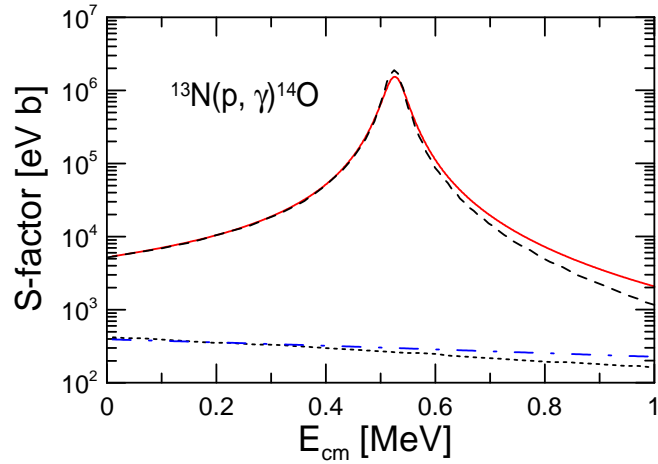}
\end{center}
\caption{(Color online). Single-particle  model calculations for the reaction $^{13}$\textrm{N}%
(\textrm{p}, $\gamma$)$^{14}$\textrm{O} are shown as a dotted-dashed
line (non-resonant) and a solid line (E1 resonance). R-matrix
results from Ref. \cite{Bing07} are also shown as a dashed line
(resonance) and a dotted
line (non-resonant).}%
\label{13Np}%
\end{figure*}

\begin{figure*}[ptb]
\begin{center}
\includegraphics[
height=2.4 in, width=3.37 in]{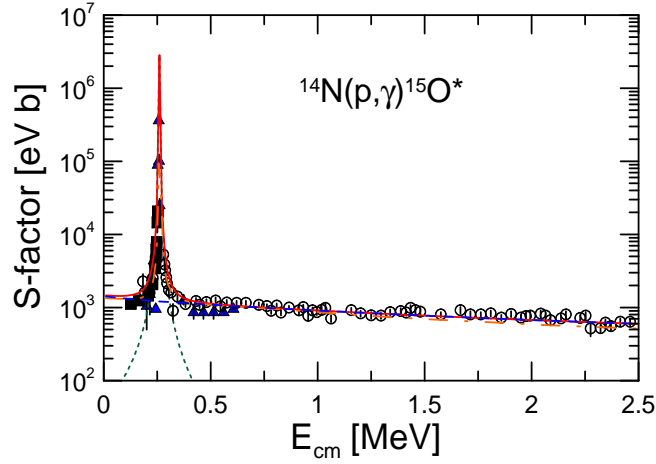}
\end{center}
\caption{(Color online). Single-particle  model calculations for $^{14}$\textrm{N}%
$({\rm p},\gamma)^{15}\text{O}$ capture to the  $6.793$ MeV excited
state of $^{15}\text{O}$. Dashed line is for the non-resonat
capture, dotted line is for the M1 resonance, and the solid line is
the total S-factor. The experimental data are from Refs.
\cite{Formicola04, Pixley57, Schroeder87}. The dotted-dashed line is
a R-matrix fit obtained in Ref. \cite{Formicola04} with the channel
radius $a=5.5$ fm (this curve is almost invisible
because it is very close to our results).}%
\label{14Np}%
\end{figure*}

\begin{figure*}[ptb]
\begin{center}
\includegraphics[
height=2.4 in, width=3.37 in ]{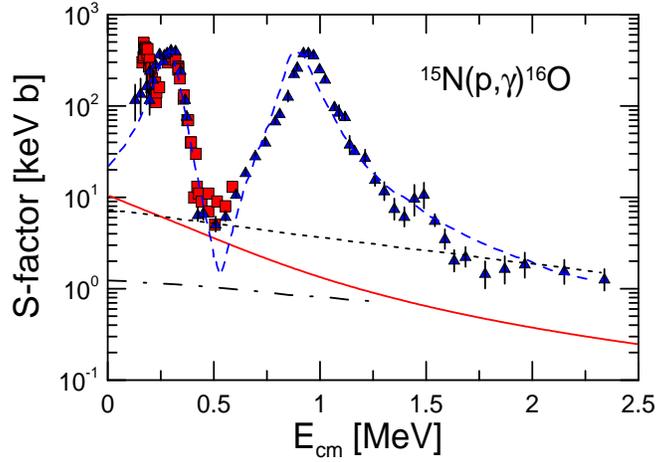}
\end{center}
\caption{(Color online). Single-particle  model calculation for the reaction $^{15}$\textrm{N}%
$({\rm p},\gamma)^{16}\text{O}$ (solid line). The experimental data
are from Refs. \cite{RR74, Hebbard60}. Dashed lines are Breit-Wigner
fits to the resonances, as described in Ref. \cite{RR74}. The dotted
line is a non-resonant capture of Ref. \cite{RR74}.  The
dotted-dashed line
represents the non-resonant capture calculation from Ref. \cite{MBB08}.}%
\label{15Np}%
\end{figure*}

\begin{figure*}[ptb]
\begin{center}
\includegraphics[
height=2.4 in, width=3.37 in ]{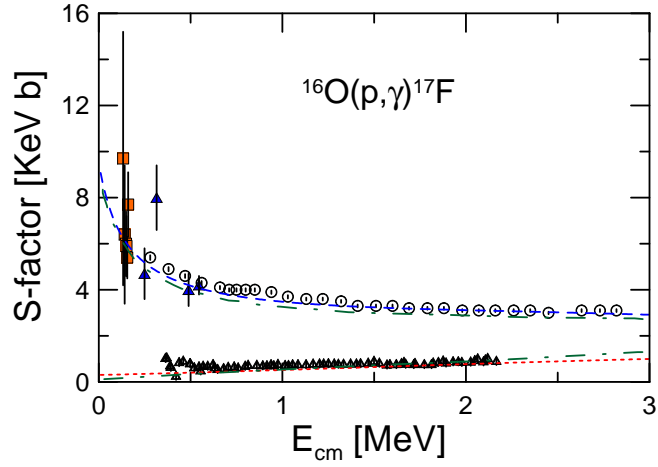}
\end{center}
\caption{(Color online). Single-particle  model calculation for the reaction $^{16}$\textrm{O}%
(p$,\gamma)^{17}\text{F}$. The dotted line and the dashed line are
for the capture to the ground state and to the first excited state
respectively. The experimental data are from Refs. \cite{HPL58,
Tanner59, Rolfs73, MKM97}. The dotted-dashed lines are the result of
shell model calculations
published in Ref. \cite{Bennaceur00}.}%
\label{16Op}%
\end{figure*}

\begin{figure*}[ptb]
\begin{center}
\includegraphics[
height=2.4 in, width=3.40 in] {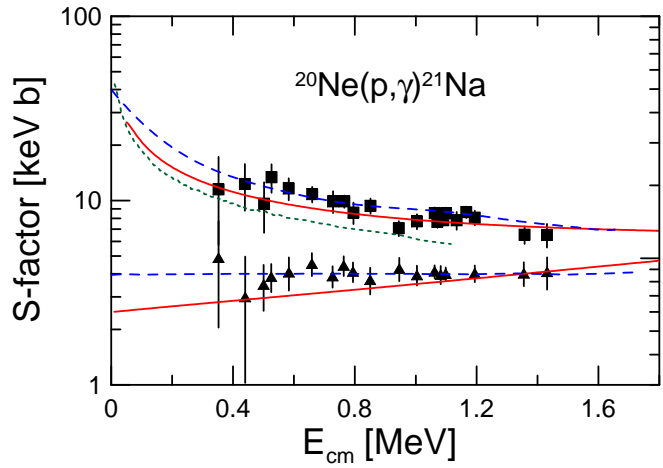}
\end{center}
\caption{(Color online). Single-particle model calculation for the reaction $^{20}$\textrm{Ne}%
(p,$\gamma)^{21}\text{Na}$. Upper solid line is for the capture to
the 2.425 MeV excited state of $^{20}$Ne and lower solid line for
the 0.332 MeV excited state. Experimental data are
from  Ref. \cite{RRSW75}. The dashed  and dotted lines are theoretical results from Ref. \cite{RRSW75} and Ref. \cite{Mukhamedzhanov06}, respectively}%
\label{20Nep}%
\end{figure*}

\begin{figure*}[ptb]
\begin{center}
\includegraphics[
height=2.4 in, width=3.37 in ]{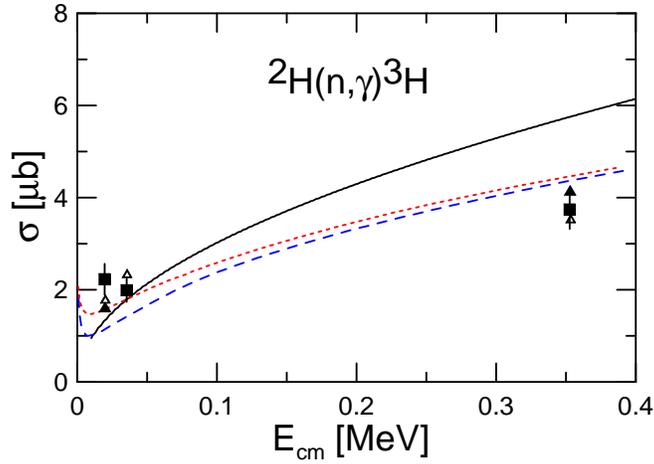}
\end{center}
\caption{(Color online). Single-particle model calculation for
$^{2}$\textrm{H}$(n,\gamma )^{3}\text{H}$ (solid line). The
experimental data are from Ref. \cite{Nagai06}. The phenomenological
results (parameter fit) from from Ref. \cite{Nagai06} are shown by
dashed  and dotted lines. Also shown are microscopic calculations
with (open trianges) and
without (solid triangles) a three-body interaction.}%
\label{dn}%
\end{figure*}

\begin{figure*}[ptb]
\begin{center}
\hspace{-0.45 cm}\includegraphics[ height=2.4 in, width=3.37
in]{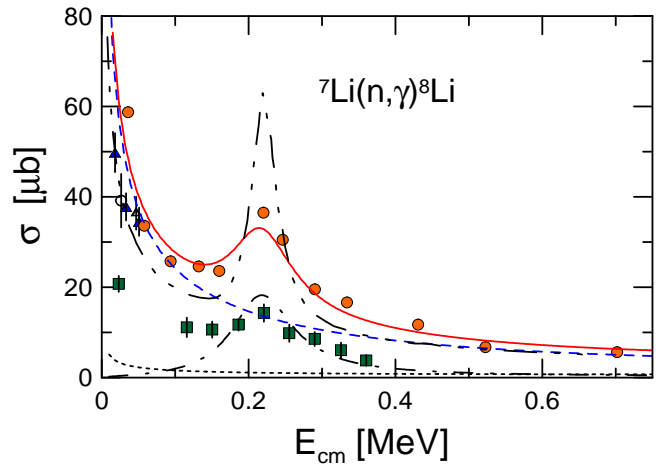}
\end{center}
\caption{(Color online). Single-particle model calculation for the
reaction $^{7}$Li(n,$\gamma $)$^{8}$Li. The dashed and dotted lines
are for the capture to the ground state and first excited state,
respectively. The dotted-dashed line is the calculated M1 resonance.
The total cross section is shown as a solid line. The calculation
result from Ref. \cite{Nagai} is shown as a dotted-dotted-dashed
line. The experimental data
are from refs. \cite{Nagai, Imhof, WSK89, Nag91, Hei98}.}%
\label{7Lin}%
\end{figure*}

\begin{figure*}[ptb]
\begin{center}
\includegraphics[
height=2.4 in, width=3.58 in]{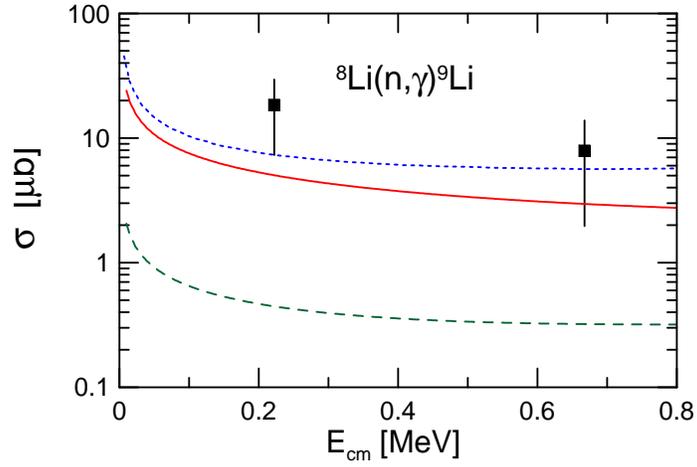}
\end{center}
\caption{(Color online). Single-particle  model calculation for
$^{8}$\textrm{Li}$(n,\gamma )^{9}\text{Li}$. The solid and the
dashed lines are the calculations for the capture to the ground and
the 1st excited states, respectively. The experimental data are from
Ref. \cite{ZGG98} using the Coulomb dissociation of $^{9}\text{Li}$
on $\text{Pb}$ targets at 28.5 MeV/A beam energy. The dotted line is
the calculation reported in Ref.
\cite{Banerjee08} for the capture to the ground state.}%
\label{8Lin}%
\end{figure*}

\begin{figure*}[ptb]
\begin{center}
\includegraphics[
height=2.4 in, width=3.37 in]{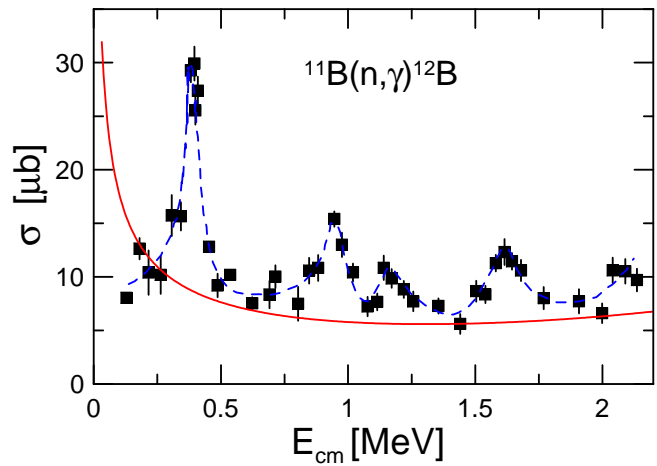}
\end{center}
\caption{(Color online). Single-particle model calculation for the (non-resonant) capture reaction $^{11}$\textrm{B}%
$(n,\gamma)^{12}\text{B}$ (solid line). The experimental data are
from Ref. \cite{IJV62}. The dashed line is a sum of fitted
Breit-Wigners superimposed to
the non-resonant capture calculation, following Ref. \cite{IJV62}.}%
\label{11Bn}%
\end{figure*}

\begin{figure*}[ptb]
\begin{center}
$%
\begin{array}
[c]{cc}%
\includegraphics[
height=2.0in, width=3.30in ]{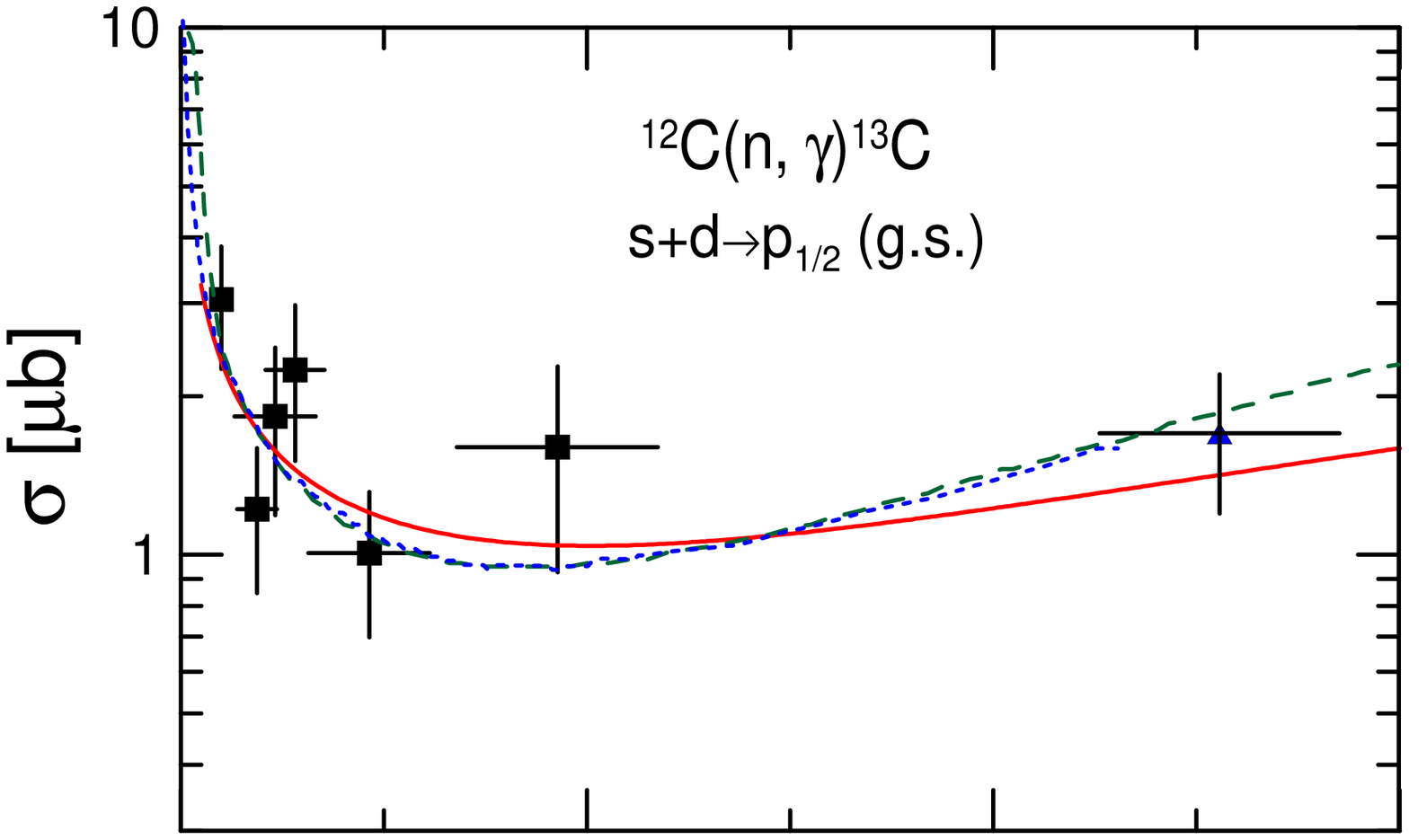} & \\
\hspace{0.1cm} \includegraphics[ height=2.4in, width=3.43in
]{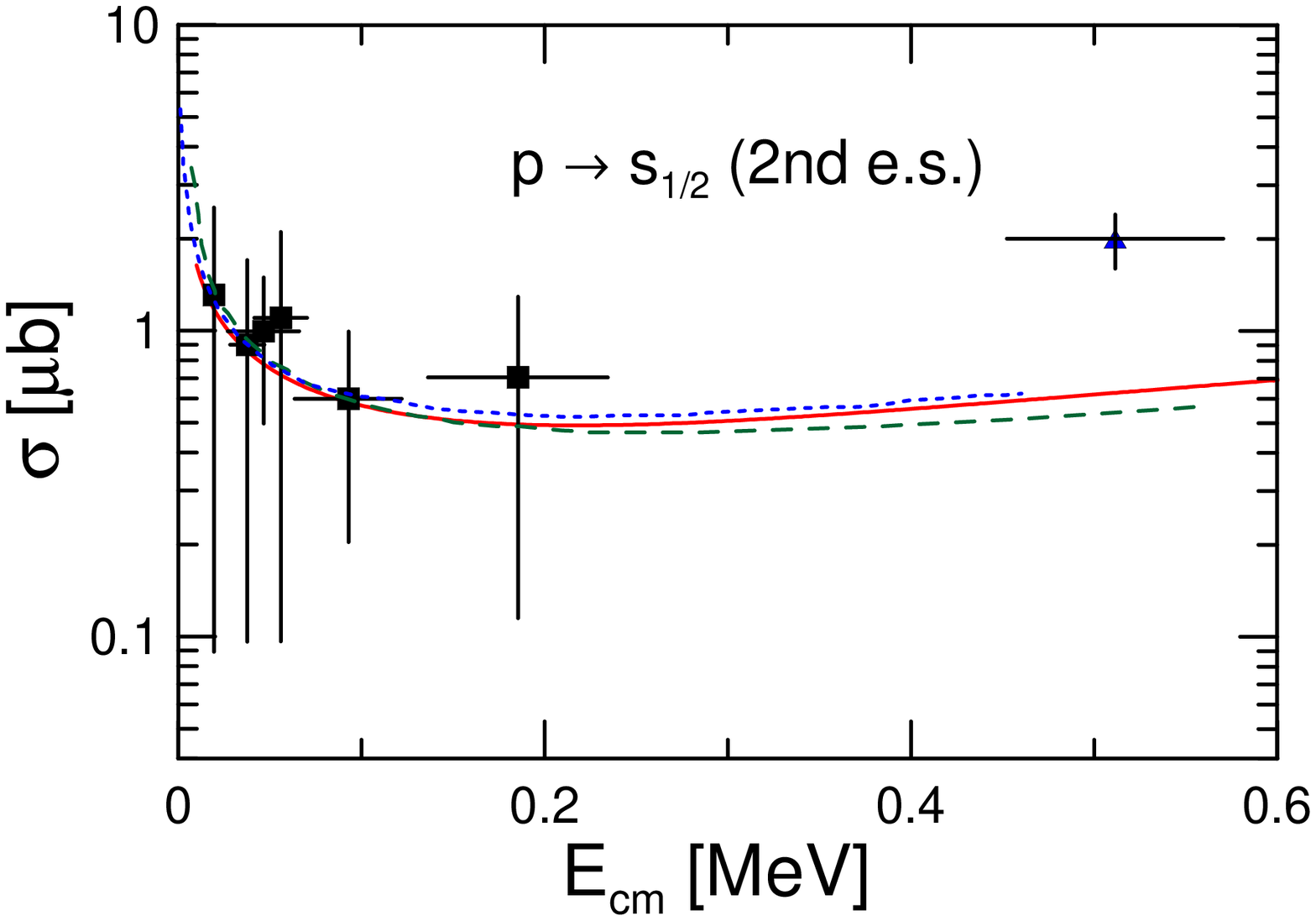} &
\end{array}
$
\end{center}
\caption{(Color online). Single-particle model calculation for
$^{12}$\textrm{C}(\textrm{n}, $\gamma$)$^{13}$\textrm{C} (solid
line).The upper panel is for the capture to the ground state whereas
the lower one is for capture to the 2nd excited state.  The
experimental data are from Ref. \cite{Ohsaki94} (filled square) and
Ref. \cite{Kikuchi95} (filled triangle).
The theoretical results from Ref. \cite{Kikuchi95} and Ref. \cite{Mengoni95} are shown by the dashed and the dotted lines, respectively.}%
\label{12Cn}%
\end{figure*}

\begin{figure*}[ptb]
\begin{center}
$%
\begin{array}
[c]{cc}%
\includegraphics[
height=2.0in, width=3.37in ]{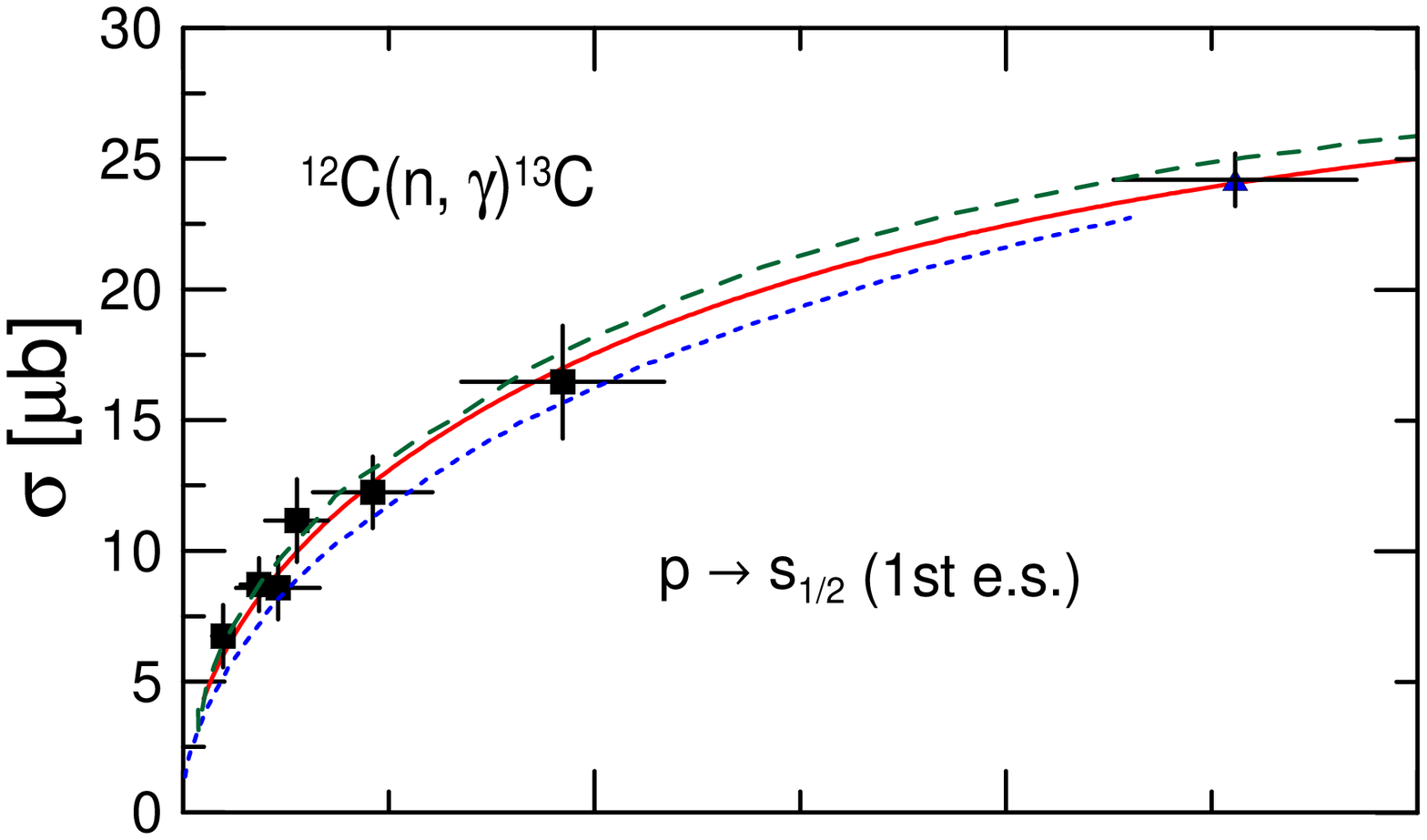} & \\ \hspace{0.4cm}
\includegraphics[ height=2.4in, width=3.37in ]{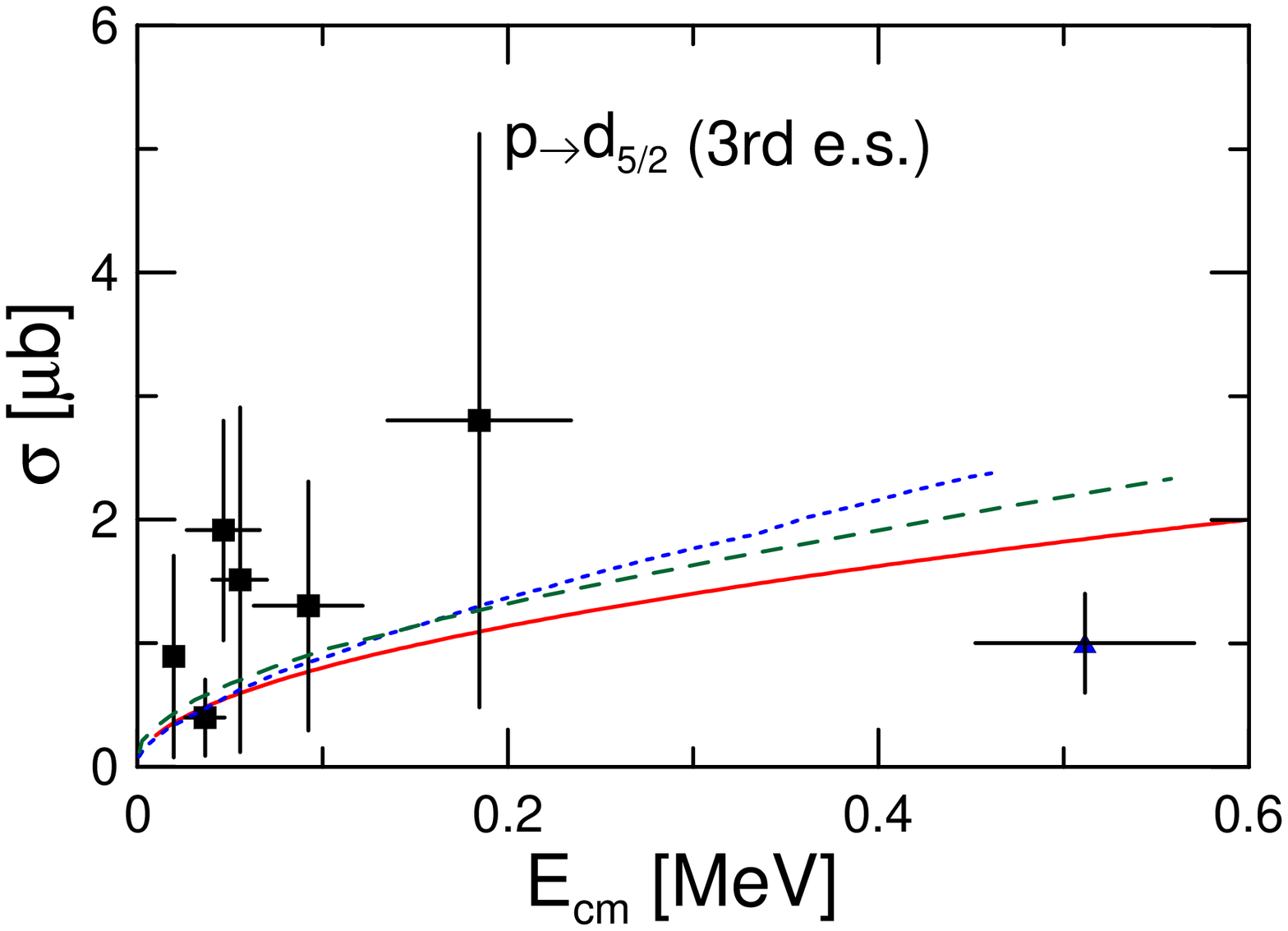} &
\end{array}
$
\end{center}
\caption{(Color online). The same as Fig. \ref{12Cn}, but for the
transitions to the 1st
excited state (upper panel) and to the 3rd excited state (lower panel).}%
\label{12Cn2}%
\end{figure*}

\begin{figure*}[ptb]
\begin{center}
\includegraphics[
height=2.4 in, width=3.37 in]{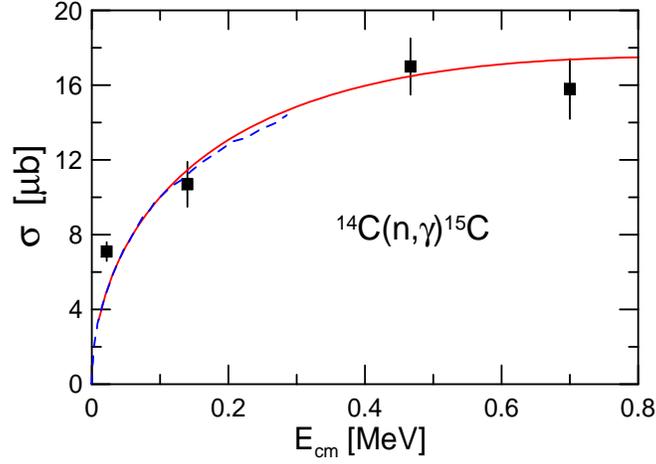}
\end{center}
\caption{(Color online). Single-particle  model calculation for the reaction $^{14}$\textrm{C}%
$(n,\gamma)^{15}\text{C}$ (solid line). The experimental data are
from Ref. \cite{Reifarth08}. The dashed line is the result from Ref. \cite{Wiescher90} using a similar potential model.}%
\label{14Cn}%
\end{figure*}

\begin{figure*}[ptb]
\begin{center}
\includegraphics[
height=2.4 in, width=3.37 in ]{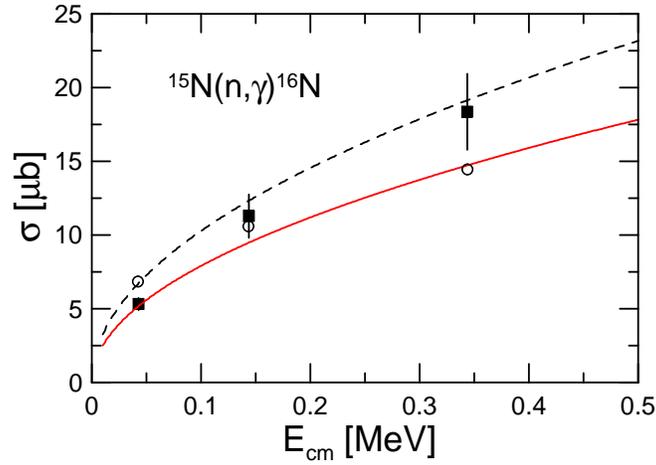}
\end{center}
\caption{(Color online). Single-particle model calculation results for $^{15}$\textrm{N}%
$(n,\gamma)^{16}$\textrm{N} (solid line). The experimental data are
from Ref. \cite{MSH96}. The non-resonant capture calculations of
Ref. \cite{MSH96} is shown by open circles. Increasing the values of
the spectroscopic values by
30\% (compatible with the experimental errors)  yields the dashed line. }%
\label{15Nn}
\end{figure*}

\begin{figure*}[ptb]
\begin{center}
$
\begin{array}
[c]{cc}
\includegraphics[height=2.4 in, width=3.37 in]{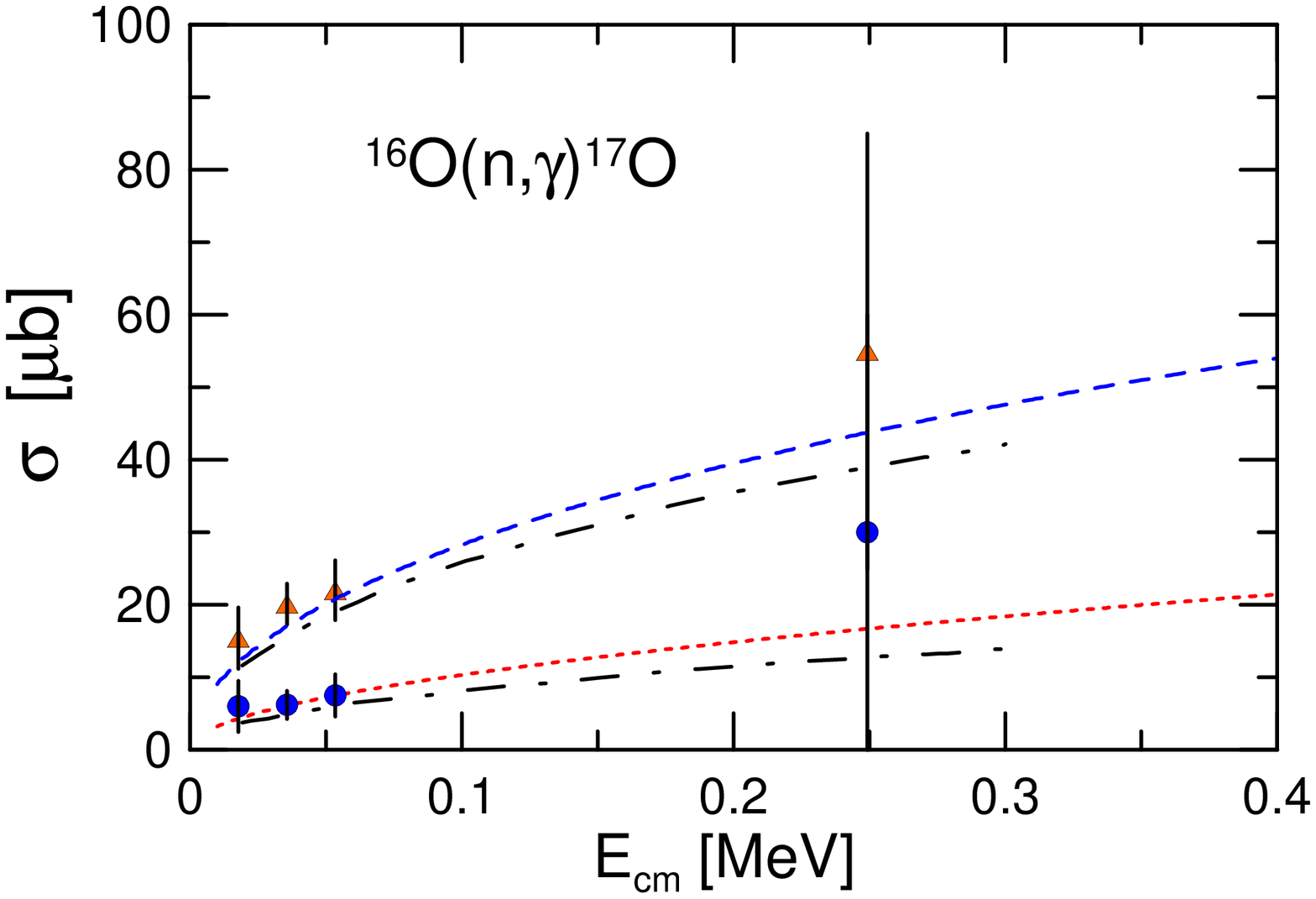} & \\
\includegraphics[height=2.4 in, width=3.37 in]{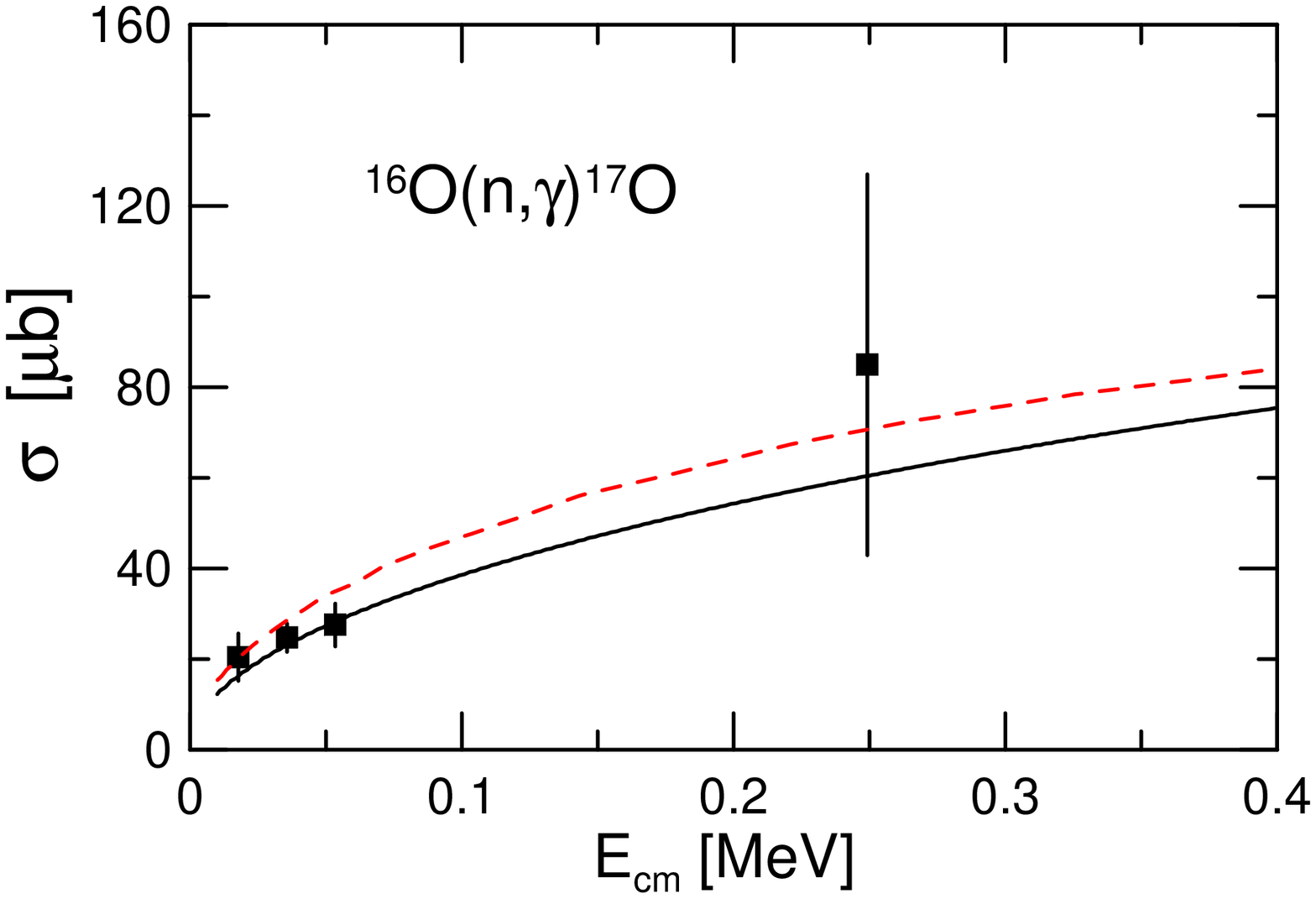} &
\end{array}
$
\end{center}
\caption{(Color online). Single-particle model calculation for
reaction $^{16}$\textrm{O} $(n,\gamma)^{17}\text{O}$ (solid lines).
The experimental data are from Ref. \cite{INM95}. Top panel: the
capture to the ground state (dotted line, filled circles) and first
excited state (dashed line, filled triangles) of $^{17}\text{O}$ are
shown separately. The results of a microscopic multicluster model
from Ref. \cite{Dufour01} are shown by dotted-dashed lines  for
comparison. Bottom panel: the total cross section of
$^{18}$\textrm{O}$(n,\gamma)^{19}\text{O}$ (solid line). The result
from Ref. \cite{CKH08} is shown as a dashed line.} \label{16On}
\end{figure*}

\begin{figure*}[ptb]
\begin{center}
\includegraphics[
height=2.4 in, width=3.37 in]{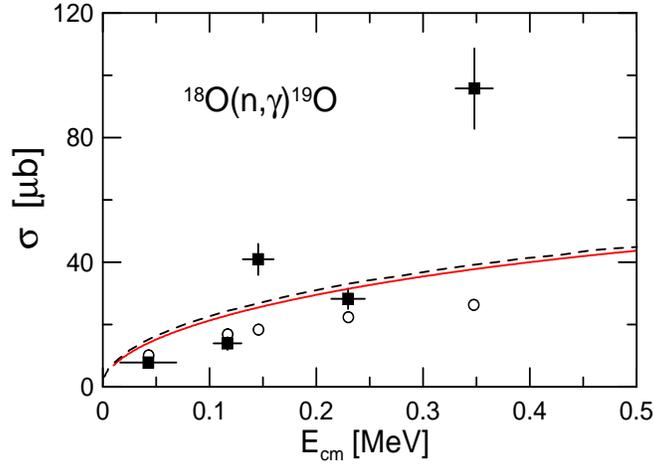}
\end{center}
\caption{(Color online). Single-particle model calculation for the reaction of $^{18}$%
\textrm{O}$(n,\gamma)^{19}\text{O}$ (solid line). The experimental
data are from Ref. \cite{Meissner96}. The non-resonant capture
calculation from Ref. \cite{Meissner96} and
\cite{Herndl99} are shown as open circles and dashed line, respectively. }%
\label{18On}%
\end{figure*}

\begin{figure*}[ptb]
\begin{center}
\includegraphics[
height=2.4 in, width=3.37 in ]{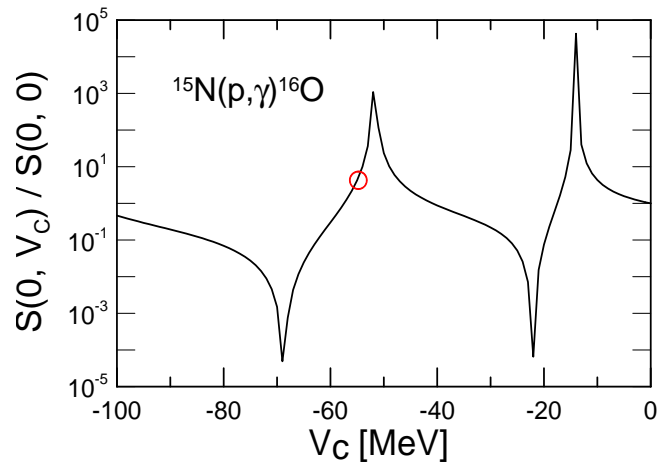}
\end{center}
\caption{(Color online). Ratio between the S-factor at $E=0$
calculated with a potential depth $V_c$ and the S-factor calculated
with a zero potential depth: $S(0,V_c)/S(0,0)$. The open circle
corresponds to the value of $V_{c}$ used in the calculation
presented in figure \ref{15Np}.}
\label{s15N}
\end{figure*}

\begin{figure*}[ptb]
\begin{center}
\vspace{0.275cm}
\includegraphics[
height=2.4 in, width=3.37 in ]{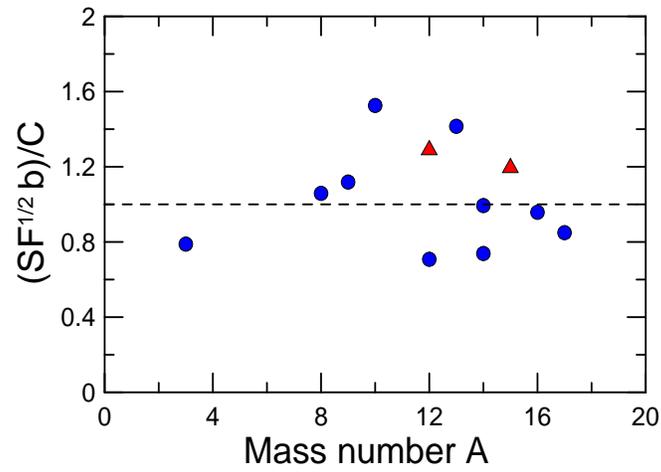} \label{Ratio}
\end{center}
\caption{(Color online). Our ANCs ($\sqrt{(SF) b^{2}}$) divided by
the ANCs obtained from references mentioned in the text as function
of the mass number $A$. The solid circles are for proton capture
whereas the solid triangles are for neutron capture. The dashed line
is equal to unity.}
\label{Ratio}
\end{figure*}

\end{document}